\documentclass{emulateapj}

\newcommand{\eps}[1]{\mbox{log~$\epsilon$\,(#1)}}

\def\kmsec{\mbox{km~s$^{\rm -1}$}}

\def\rpro{\mbox{$r$-process}}
\def\spro{\mbox{$s$-process}}
\def\ncap{\mbox{$n$-capture}}
\def\msun{$M_{\odot}$}

\shorttitle{The Ubiquity of the $r$-process}
\shortauthors{Roederer et al.}
\submitted{Accepted for publication in the Astrophysical Journal}

\begin{document}

\title{The Ubiquity of the Rapid Neutron-Capture Process\footnote{
This paper includes data taken at The McDonald Observatory of The 
University of Texas at Austin.
} }

\author{
Ian U.\ Roederer,\altaffilmark{1,2}
John J.\ Cowan,\altaffilmark{3}
Amanda I.\ Karakas,\altaffilmark{4,5}
Karl-Ludwig Kratz,\altaffilmark{6} \\
Maria Lugaro,\altaffilmark{5}
Jennifer Simmerer,\altaffilmark{7} 
Khalil Farouqi,\altaffilmark{8} and
Christopher Sneden\altaffilmark{1}
}

\altaffiltext{1}{Department of Astronomy, University of Texas at Austin,
1 University Station, C1400, Austin, TX 78712-0259 USA}

\altaffiltext{2}{Present address: Carnegie Observatories,
813 Santa Barbara Street, Pasadena, CA 91101 USA;
iur@obs.carnegiescience.edu}

\altaffiltext{3}{Homer L.\ Dodge Department of Physics and Astronomy, 
University of Oklahoma, Norman, OK 73019 USA}

\altaffiltext{4}{Research School of Astronomy \& Astrophysics, 
The Australian National University, 
Mount Stromlo Observatory, Cotter Road, Weston, ACT 2611, Australia}

\altaffiltext{5}{Centre for Stellar and Planetary Astrophysics,
School of Mathematical Sciences, Monash University, Clayton, VIC 3800,
Australia}

\altaffiltext{6}{Max-Planck-Institut f\"{u}r Chemie, Otto-Hahn-Institut,
J.-J.-Becherweg 27, D-55128 Mainz, Germany}

\altaffiltext{7}{Department of Physics and Astronomy, University of Utah,
115 S.\ 1400 E., Salt Lake City, UT 84112-0830 USA}

\altaffiltext{8}{Zentrum f\"{u}r Astronomie der Universit\"{a}t
Heidelberg, Landessternwarte, K\"{o}nigstuhl 12, D-69117 Heidelberg,
Germany}

\begin{abstract}
To better characterize the abundance patterns produced 
by the $r$-process, 
we have derived new abundances or upper limits for the heavy elements
zinc (Zn, $Z =$~30),
yttrium (Y, $Z =$~39), 
lanthanum (La, $Z =$~57),
europium (Eu, $Z =$~63), and 
lead (Pb, $Z =$~82).
Our sample of 161 metal-poor stars includes new measurements from 88 
high resolution and high signal-to-noise spectra obtained with the 
Tull Spectrograph on the 2.7~m Smith Telescope at McDonald Observatory,
and other abundances are adopted from the literature.
We use models of the $s$-process in
AGB stars to characterize the high Pb/Eu ratios produced in the 
$s$-process at low metallicity,
and our new observations then allow us to identify a sample
of stars with no detectable $s$-process material.
In these stars, we find no significant 
increase in the Pb/Eu ratios with increasing metallicity.
This suggests that $s$-process material 
was not widely dispersed until the overall Galactic metallicity
grew considerably, perhaps even as high as [Fe/H]~$= -$1.4,
in contrast with earlier studies that suggested a much lower
mean metallicity.
We identify a dispersion of at least 
0.5~dex in [La/Eu] in metal-poor stars with [Eu/Fe]~$< +$0.6
attributable to
the $r$-process, suggesting that there is no unique
``pure'' $r$-process elemental ratio among 
pairs of rare earth elements.
We confirm earlier detections of
an anti-correlation between Y/Eu and Eu/Fe
bookended by stars strongly enriched in the $r$-process
(e.g., \mbox{CS~22892--052}) and those with deficiencies
of the heavy elements (e.g., \mbox{HD~122563}).
We can reproduce the range of Y/Eu ratios using simulations of
high-entropy neutrino winds of core-collapse supernovae
that include charged-particle and neutron-capture 
components of $r$-process nucleosynthesis.
The heavy element abundance patterns in most metal-poor stars
do not resemble that of \mbox{CS~22892--052},
but the presence of heavy elements
such as Ba in nearly all metal-poor stars 
without $s$-process enrichment
suggests that the $r$-process is a common phenomenon.
\end{abstract}


\section{Introduction}
\label{intro}

How much diversity exists among the heavy element abundance
patterns observed in stars?
Two general cases of nucleosynthesis, neutron ($n$) capture
on slow ($s$) or rapid ($r$) timescales relative to the 
average $\beta$-decay rates, produce clearly distinct 
abundance patterns because these processes flow through
different sets of nuclei
(e.g., \citealt{burbidge57}, \citealt{cameron57}).
Yet the overwhelming majority of present-day stars have 
been enriched by the products of multiple nucleosynthetic
events, complicating the process of disentangling the
products of individual (classes of) events on
observational grounds alone.
Theoretical work that incorporates large amounts of 
experimental nuclear input data, when available, has proved illuminating,
particularly with regard to the relative contributions of 
the $s$- and \rpro\ to Solar system (S.S.) material
(e.g., \citealt{cameron73}, \citealt{kappeler89}, \citealt{arlandini99}).
Yet neither process produces an identical 
set of nuclei in each event---variations in the physical
conditions present at the time of nucleosynthesis, availability
of seed nuclei, and the duration of the event surely conspire to 
affect the nucleosynthetic yields, whether in subtle or extreme fashion.
From this perspective, 
the discovery of metal-poor stars with a wide variety 
of \ncap\ abundance patterns in the last 20 years or so has
created a rich setting to test and refine our
understanding of the
diverse and often exotic physical conditions of 
heavy element nucleosynthesis.

\mbox{CS~22892--052},
an extremely metal-poor K~giant 
star from the HK Survey of \citet{beers92}, was identified 
by \citet{sneden94} as having a 
strong overabundance of the \ncap\ elements relative to Fe.  
The enrichment pattern could not be fit by any published 
predictions for the \spro\ process, and \citet{cowan95} showed that 
the abundance pattern from barium (Ba, $Z =$~56) to erbium 
(Er, $Z =$~68) was ``strikingly similar'' to the S.S.\ \rpro\
residuals predicted by \citet{kappeler89}.
\citet{sneden96} extended this sequence to 
thulium (Tm, $Z =$~69), 
ytterbium (Yb, $Z =$~70), 
hafnium (Hf, $Z =$~72), 
osmium (Os, $Z =$~76), and the radioactive element
thorium (Th, $Z =$~90), which can only be produced in the \rpro.
Over the last decade, several other metal-poor stars have 
been identified---including
several from the first study of \ncap\ elements in a large sample
of metal-poor stars by \citet{gilroy88}---as 
standard templates to characterize the \rpro\ 
nucleosynthesis pattern
(\mbox{HD~115444}, \citealt{westin00};
\mbox{CS~31082--001}, \citealt{hill02};
\mbox{BD$+$17~3248}, \citealt{cowan02};
\mbox{HD~221170}, \citealt{ivans06}).

The match between the stellar \rpro\ abundances and 
the scaled S.S.\ \rpro\ pattern does not always extend to the lighter
heavy elements, including strontium (Sr, $Z =$~38), 
yttrium (Y, $Z =$~39), and zirconium (Zr, $Z =$~40).
Observational evidence demanding an additional 
nucleosynthesis site for the $A <$~130 nuclei
was first presented by \citet{wasserburg96} in their analysis
of radioactive isotopes in the S.S.
This result has been expanded upon by
observations of $Z \geq$~38 elements in metal-poor stars by
numerous investigators, including
\citet{mcwilliam98}, \citet{burris00,burris09},
\citet{johnson02b}, \citet{aoki05}, \citet{barklem05},
\citet{francois07}, \citet{cohen08}, \citet{lai08}, and
\citet{mashonkina08}.

It is also apparent that some very low
metallicity stars have heavy element abundance patterns that cannot be 
matched by either the scaled S.S. \rpro\ or \spro\ components.
Following similar reasoning employed by \citet{sneden83} 
and \citet{sneden85} 
when comparing \mbox{HD~122563} and \mbox{HD~110184},
this point was made emphatically by \citet{honda06,honda07}
when comparing the heavy elements in
\mbox{HD~122563} and \mbox{HD~88609} to \mbox{CS~22892--052}.
When these stars' heavy element abundances were subtracted from
the S.S.\ \rpro\ abundance pattern, two distinct patterns emerged,
and that of \mbox{HD~122563} and \mbox{HD~88609} was incompatible
with any combination of scaled S.S.\ \rpro\ or \spro\ components
(\citealt{honda07}, their Figure~5).

We have noticed a possible anti-correlation
between the ratio of two elements in the rare earth element (REE)
domain, lanthanum (La, $Z =$~57) and europium (Eu, $Z =$~63),
and the bulk enrichment of Eu relative to Fe.
The three standards with the lowest [Eu/Fe] ratios
(\mbox{BD$+$17~3248}, \mbox{HD~221170}, and \mbox{HD~115444};
$\langle$[Eu/Fe]$\rangle = +0.8$)\footnote{
We adopt the standard spectroscopic notations that
[A/B]~$\equiv$ log$_{10}$(N$_{\rm A}$/N$_{\rm B}$)$_{\star}$~--
log$_{10}$(N$_{\rm A}$/N$_{\rm B}$)$_{\odot}$ and
log~$\epsilon$(A)~$\equiv$ 
log$_{10}$(N$_{\rm A}$/N$_{\rm H}$)~$+$~12.00
for elements A and B.} 
have log~(La/Eu)~$= +0.21 \pm 0.06$ \citep{sneden09}, while the 
two standards with the highest [Eu/Fe] ratios
(\mbox{CS~22892--052} and \mbox{CS~31082--001};
$\langle$[Eu/Fe]$\rangle = +1.6$) have
log~(La/Eu)~$= +0.10 \pm 0.01$ \citep{sneden09}.
The star with the highest level of \rpro\ enrichment known
(\mbox{HE~1523--0901}, [Eu/Fe]~$=+1.8$; \citealt{frebel07}),
has log~(La/Eu)~$= -0.01$.

Here we systematically examine the 
relationship between the light (e.g., Y) and heavy (e.g., La, Eu, and Pb)
abundances in these stars and others to better characterize
the abundance patterns observed in metal-poor stars
and illuminate the nature of the 
nucleosynthetic process(es) that might be responsible for
producing them.
Sections~\ref{sample} and \ref{nos} 
describe our sample, new abundance derivations, and
attempts to identify any trace of \spro\ material in these stars.
Section~\ref{correlations} describes the observed correlation
between the light and heavy \ncap\ elements for the $r$-only stars, and 
Section~\ref{hew} describes a plausible physical model
to explain this correlation.
Finally, in Sections~\ref{discussion} and \ref{conclusions}, we
discuss the implications of this result and summarize our findings.

\section{Sample and Abundance Analysis}
\label{sample}

\citet{simmerer04} obtained high resolution ($R \sim$~60,000)
and high signal-to-noise (S/N~$\sim$~100 at 4100\AA) spectra
for 88 bright ($V \leq$~11.0) metal-poor dwarf and giant 
stars from the halo and disc
using the Tull Cross-dispersed Echelle Spectrograph
\citep{tull95} on the 2.7m Smith Telescope at McDonald Observatory.
We adopt the atmospheric parameters from \citet{simmerer04} and
derive new zinc (Zn, $Z =$~30), Y, and Pb
abundances for the stars in this sample.\footnote{
We exclude HD~232078, which has an effective temperature
more than 200~K cooler than any other star in the sample
($T_{\rm eff} =$~3875~K).}
Abundances are derived using the current version of the 
spectral analysis code MOOG
\citep{sneden73}, assuming that all lines are formed 
under conditions of local thermodynamic equilibrium in a
one dimensional, plane-parallel atmosphere.

\begin{figure*}
\epsscale{1.00}
\begin{center}
\plottwo{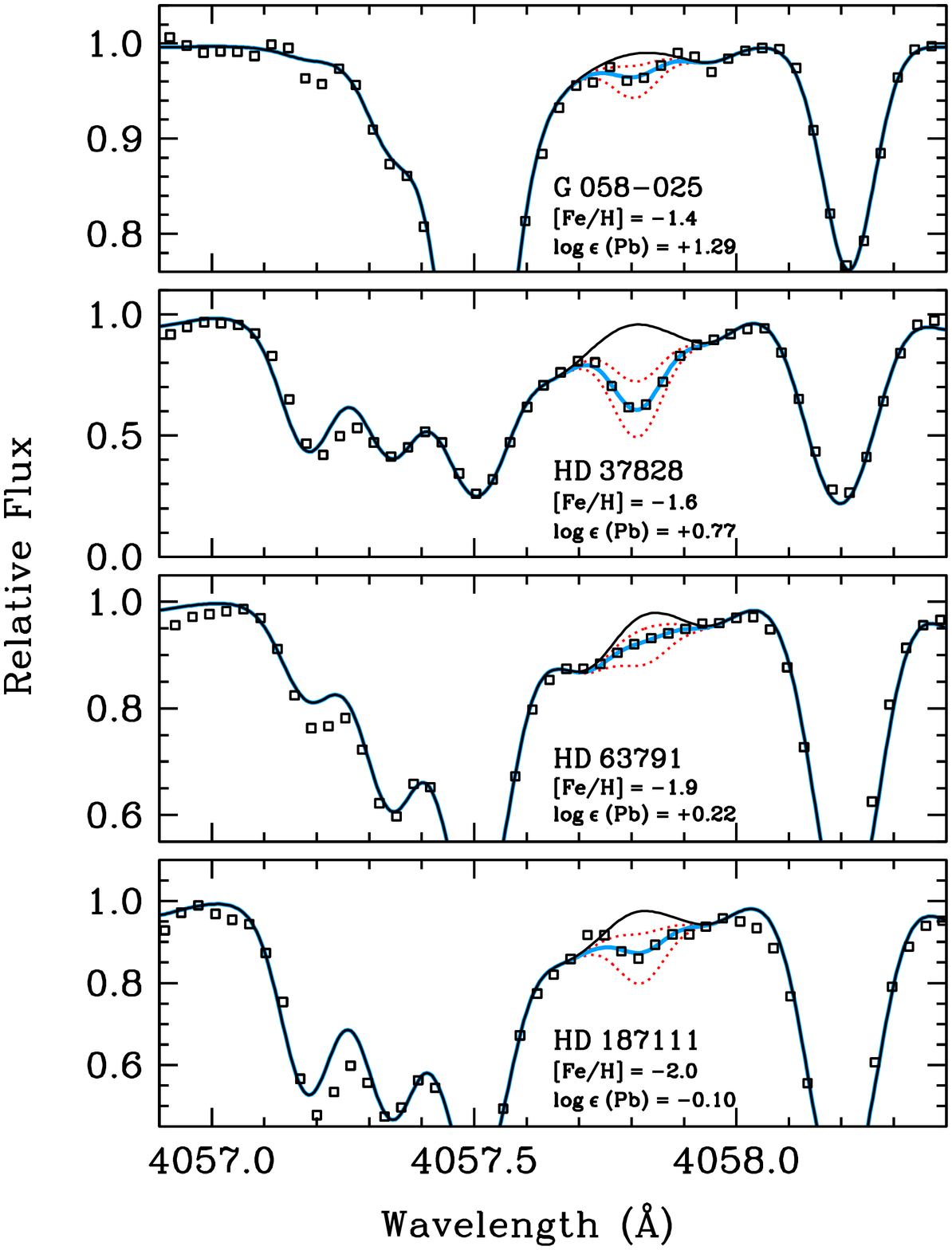}{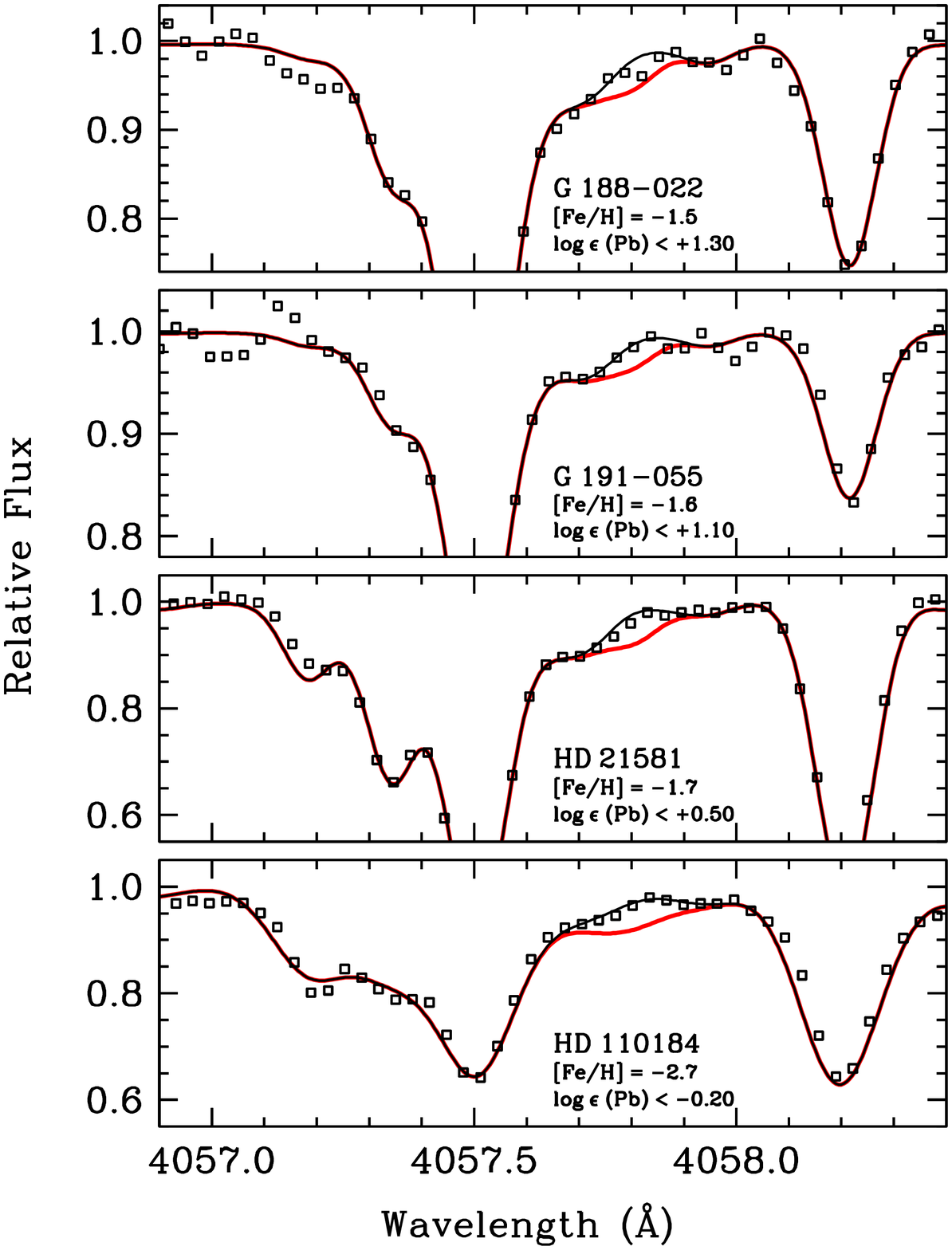}
\end{center}
\caption{
\label{pbspecplot}
Observed and synthetic spectra of the Pb~\textsc{i} region
in 8~stars.
The 4~panels on the left show stars with Pb detections, while
the 4~panels on the right show stars with no Pb detection.
On the left, the bold line indicates our best-fit synthesis,
the dotted lines indicate $\pm$~0.30~dex from this abundance,
and the thin line indicates a synthesis with no Pb present.
On the right, the bold line indicates our Pb upper limit
and the thin line indicates a synthesis with no Pb present.
The observed spectrum is indicated by the open squares.
}
\end{figure*}

Zn is the heaviest element in the Fe-group that is readily
accessible in the optical regime, and we use the
Zn~\textsc{i} 4722 and 4810\AA\ lines as abundance indicators.
The Sr~\textsc{ii} resonance lines at 4077 and 4215\AA\ are 
saturated or blended in most of these stars, so we instead
derive abundances for the next heavier element, Y, 
using the Y~\textsc{ii} lines at 4883, 5087, and 5200\AA.
Equivalent widths for these lines are measured within
the IRAF environment,\footnote{
IRAF is distributed by the National Optical Astronomy Observatories,
which are operated by the Association of Universities for Research
in Astronomy, Inc., under cooperative agreement with the National
Science Foundation.}
and these equivalent widths are reported in Table~\ref{ewtab}.
Abundances of Zn~\textsc{i} and Y~\textsc{ii} are derived 
by requiring that the predicted line-by-line abundances fit the
measured equivalent widths and then averaging the abundance over all lines.
We adopt the log($gf$) values for Zn~\textsc{i} and Y~\textsc{ii} 
from \citet{biemont80} and \citet{hannaford82}, respectively,
which are routinely employed in studies of metal-poor stars and
were found by \citet{biemont80} and \citet{hannaford82} to
yield reliable abundances for lines in the Solar photosphere.
The Pb~\textsc{i} abundance was derived from the 4057\AA\ line by
fitting synthetic spectra to match the observed spectrum.
This line is often weak and nearly always blended in our spectra.
When the Pb~\textsc{i} line cannot be detected, we derive an
upper limit on its abundance.
Several examples of our fits and upper limits are presented in
Figure~\ref{pbspecplot}.
We adopt the Pb~\textsc{i} log($gf$) values of \citet{biemont00},
which is also the most commonly-used source for these data.
No additional broadening of the 4057\AA\ line,
caused by isotope shifts or hyperfine structure of the $^{207}$Pb isotope,
could be detected.

Final abundances for Zn~\textsc{i}, Y~\textsc{ii}, and Pb~\textsc{i} 
are reported in Table~\ref{abundtab} along with the
[Fe/H], La~\textsc{ii}, and Eu~\textsc{ii} abundances
derived by \citet{simmerer04} (whose study was limited to C, Fe,
La, and Eu).
We have supplemented this sample with metal-poor stars from other 
recent studies.
These abundances, along with the original source references,
are summarized in Table~\ref{abundtab}.
We have not made any explicit corrections to the abundances to 
put them on a common log($gf$) scale, but the laboratory sources
for the five species examined here are commonly used, 
and all predate the abundance measurements compiled here.

\section{Identifying Stars with No $s$-process Material}
\label{nos}

Nucleosynthesis products of the \rpro\ generally are visible in
the lowest metallicity stars with detectable heavy elements,
and products of the \spro\ typically appear in higher metallicity
stars that were formed later (e.g., \citealt{gratton94},
\citealt{burris00}, \citealt{simmerer04}).
\citet{cowan96}, for example, noted that the heavy element abundance
pattern in \mbox{HD~126238} ([Fe/H]~$= -$1.7)
could be fit by assuming a majority contribution
from the scaled S.S.\ \rpro\ and a small fraction of the
total S.S.\ \spro\ abundance.
The \spro\ contribution was necessary to account for the
slight overabundances (relative to the scaled S.S.\ \rpro\ 
pattern normalized at Eu) of Ba--Nd ($Z =$~56--60) and Pb.
To assess whether this abundance pattern
may actually result from repeatable and quantifiable dispersion 
in the \rpro\ itself, we need to remove from our sample
all stars with even the slightest hint of \spro\ material.
We outline here several approaches to identify these stars.

The \spro\ occurs in the deep He-rich layer of stars on the asymptotic
giant branch (AGB)
and \spro\ products are carried to the envelope via dredge-up episodes 
(the third dredge-up) 
and shed into the interstellar medium (ISM) via strong winds. 
The $^{22}$Ne($\alpha$,n)$^{25}$Mg and 
$^{13}$C($\alpha$,n)$^{16}$O reactions provide the neutrons for the \spro.
The former is activated in the convective regions 
that develop episodically in connection with partial 
He burning (thermal pulses), 
while the latter is activated during the interpulse periods 
(see \citealt{busso99} for a review). 
When compared with an \spro\ operating in a metal-rich environment,
at low metallicity the \spro\ produces large Pb/Fe 
(and, e.g., Pb/Eu, Pb/Ba, and Pb/Sr) ratios.
(see, e.g., \citealt{gallino98} and Section~\ref{pbmodels} below).
Thus, enhanced Pb/Fe and Pb/Eu ratios should be clear indicators
of low-metallicity \spro\ nucleosynthesis.
This phenomenon is gradually muted with increasing metallicity
of the \spro\ environment,
reaching a maximum efficiency of Pb production 
around [Fe/H]~$\sim -$1.0 \citep{travaglio01}.
For the present study, to minimize our dependence on any particular
set of AGB \spro\ models, we conservatively assume that high Pb/Fe and Pb/Eu
ratios are only obtained in environments with [Fe/H]~$< -$1.4
(cf.\ \citealt{bisterzo10}).

\begin{figure*}
\epsscale{0.85}
\begin{center}
\plotone{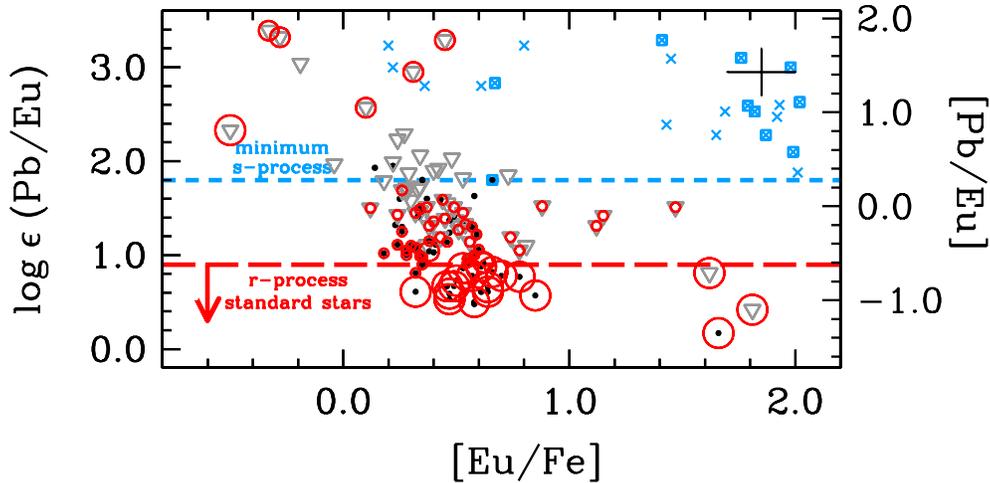}
\end{center}
\caption{
\label{pbeuplot}
Logarithmic Pb/Eu ratios as a function of [Eu/Fe].
All upper limits are indicated by downward-facing triangles, 
and all measurements are indicated by small black circles.
All red circles represent stars lacking any detectable
trace of \spro\ material, and
the relative size of the circles identifies the method
we have used to characterize them
(large circles: 
stars with [Pb/Eu]~$\leq -0.6$ as well as 
\mbox{HD~88609} and \mbox{HD~122563};
medium circles:
members of the stellar stream analyzed by \citealt{roederer10a};
small circles: 
stars with [Pb/Eu]~$\leq +0.3$
and [Fe/H]~$< -$1.4).
The long-dashed line indicates 
[Pb/Eu]~$\leq -0.6$
(the upper extent
of the range of Pb/Eu for the \rpro\ standard stars),
and the short-dashed line indicates [Pb/Eu]~$\leq +0.3$
(the approximate minimum ratio expected from AGB pollution).
For comparison, 
small blue ``X''s denote stars enriched in \spro\ material,
and small open squares around these ``X''s indicate that
the star shows RV variations.
A representative uncertainty is shown 
in the top right corner.
}
\end{figure*}

The handful of \rpro\ standard stars with [Fe/H]~$\sim -$3.0
show Pb abundances or upper limits consistent with the low
levels expected if no \spro\ material is present;
these Pb measurements are also consistent with or slightly lower than
(e.g., \mbox{CS~31082--001} and \mbox{HE~1523$-$0901};
\citealt{plez04}, \citealt{frebel07})
\rpro\ model predictions \citep{kratz04,roederer09b}.
These stars all have [Pb/Eu]~$\leq -$0.8 or $-$0.7
(log~(Pb/Eu)~$\leq +$0.7 or $+$0.8).
From this evidence, we conclude that all stars with [Fe/H]~$< -$1.4 and 
[Pb/Eu]~$\leq -$0.6
(log~(Pb/Eu)~$\leq +$0.9) contain no \spro\ material.
This low level of Pb is perhaps the best diagnostic for selecting
metal-poor stars containing no material produced by the \spro.
This limit is illustrated in Figure~\ref{pbeuplot}.
This figure indicates that a number of metal-poor stars can 
be diagnosed as $r$-only using this criterion alone.

\citet{honda06,honda07} have performed extensive studies of 
the heavy elements in \mbox{HD~88609} and \mbox{HD~122563}.
They concluded
that the enrichment patterns in these stars cannot be fit by 
the scaled S.S.\ \rpro\ pattern, abundances predicted by \spro\ models,
or any combination of these
(see also \citealt{sneden83}, \citealt{farouqi08}, and \citealt{kratz08a}).
No elements heavier than the REE group have been detected 
in these stars (including Pb), 
and we likewise assume that they contain no \spro\ material.

Finally, a variation of this principle was used by \citet{roederer10a} to 
deduce that no \spro\ enrichment had occurred in a metal-poor
stellar stream.
These stars' similar kinematics imply that they originated
in a common (but unknown)
progenitor system that may have been shredded by the Milky Way.
The \ncap\ elements exhibited a range of X/Fe ratios, but the 
\ncap\ abundance pattern (e.g., X/Eu) was itself unchanged in all
stream members and matched the scaled abundance pattern 
of the \rpro\ standard star \mbox{CS~22892--052} for the
heavy \ncap\ elements.
Pb could only be detected in the two most metal-rich stars in the stream
([Fe/H]~$= -$1.5 and $-$1.6), but in these two cases the Pb 
abundance was low and consistent with the Pb/Eu ratio expected 
for enrichment by the \rpro.
If the \spro\ had not enriched the most metal-rich stars in the stream,
it is highly unlikely that it enriched the more metal-poor stars.
Since all stream members show the same general \ncap\ abundance pattern,
we contend that all stars studied in this particular stream show no evidence
of \spro\ material.

\subsection{Low Metallicity Models of AGB $s$-process Nucleosynthesis}
\label{pbmodels}

To further investigate the minimum [Pb/Eu] and [La/Eu]
ratios that may be produced in the \spro,
we have computed AGB nucleosynthesis models for a range of 
stellar masses at metallicities of [Fe/H]~$= -$1.4 and $-$2.3.
We use techniques 
described in \citet{karakas09} but with an extended network of 
291 species from H to S and Fe to Bi, assuming a 
scaled-solar initial composition and using reaction rates taken from the 
JINA REACLIB database \citep{cyburt10}.
We refer to \citet{karakas07} and \citet{karakas10} for a full 
description of the stellar structure models and input parameters. 
The resulting [La/Eu] and [Pb/Eu] ratios, determined
at the end of the AGB phase, are shown in Table~\ref{agbtab}.
The low-mass AGB models ($M \lesssim 3.5M_{\odot}$) show surface 
compositions of [Pb/Eu]~$\ge +1.6$. 
In contrast, intermediate-mass AGB models ($M \gtrsim 4M_{\odot}$) 
predict lower ratios, [Pb/Eu]~$\gtrsim +0.3$. 
Hence, the minimum values produced by our models are
[La/Eu]~$= +$0.61 (log~(La/Eu)~$\approx +$1.2) and
[Pb/Eu]~$= +$0.37 (log~(Pb/Eu)~$\approx +$1.9). 

The result of high Pb/Fe and Pb/Eu ratios is generally model-independent
because the \spro\ at metallicities below [Fe/H]~$\lesssim -1$ 
favors production of Pb over lighter \ncap\ elements.
This follows from the fact that the $^{13}$C neutron source is primary,
formed from the H and He initially present in the star.
The Fe-group seed nuclei for the \spro\ are not primary elements.
Thus, the time-integrated neutron flux is proportional to 
$^{13}$C/$Z$, and at lower metallicity ($Z$) the neutron exposure
increases favoring the production of heavier elements in the \spro\
\citep{clayton88,gallino98}.
This enhanced Pb phenomenon
is evident in the predictions of our models shown in Table~\ref{agbtab}.

The more massive AGB stars are relatively short lived 
($\lesssim 100$~Myr) and therefore could have contributed 
early \spro\ enrichment of the halo.
In a standard initial mass function they constitute only a 
few percent of all AGB stars.
Because of their relatively quick evolution they may be
more likely to have injected \spro\ enriched material into the ISM
from which our low-metallicity stellar sample formed,
so we include their \spro\ yields when considering the lowest
[Pb/Eu] ratios that may be produced.
Given sufficient time, however, their contributions will be diluted by
those from the lower-mass stars that produce higher [Pb/Eu] ratios.

The main uncertainty in AGB \spro\ predictions is the formation 
of the main neutron source nucleus $^{13}$C. 
In order to have enough $^{13}$C for the \spro\ to occur, extra 
mixing is needed to carry protons from the convective envelope down into the 
$^4$He- and $^{12}$C-rich radiative layer of the star.
This typically occurs when a sharp discontinuity between 
these two regions is left after the the third dredge-up. 
These protons can then react with $^{12}$C to produce a region rich in 
$^{13}$C and $^{14}$N (the $^{13}$C ``pocket''). 
The physical mechanism leading to this mixing is not known, and thus 
its dependence on the stellar mass and metallicity is also unknown.
In the stellar models with $M >$~3~\msun\ we do not include a 
$^{13}$C pocket.
It has been qualitatively shown that in this mass and metallicity range
protons mixed down from the envelope
into the deeper layers burn while being mixed. 
The detailed consequences of proton ingestion on the nucleosynthesis 
are not well known
but could range from the inhibition of formation of the 
$^{13}$C pocket \citep{goriely04} 
to termination of the AGB phase altogether \citep{woodward08}. 
The intermediate-mass models of low-metallicity of 
\citet{herwig04} show the formation of a $^{13}$C pocket; however, 
this occurs deep in the star below the He shell, where there 
is very little $^{4}$He. 
In summary, the $^{13}$C neutron source is most likely not 
available or not efficient in these stars. 

For the lower mass AGB stars we treat the formation of the 
$^{13}$C pocket in an artificial way 
as described in detail in \citet{lugaro04}.
At the end of each third dredge-up episode
we add an exponentially decaying proton profile 
from the envelope value $\simeq$~0.7 
to 10$^{-4}$ at a point in mass 0.002~\msun\ 
below the base of the envelope
in the $^{12}$C-rich layer.
This choice results in an \spro\ rich region of $\simeq$~0.001~\msun\ 
because the \spro\ occurs only in the bottom half 
of the resulting $^{13}$C pocket 
where there are fewer $^{14}$N atoms to capture neutrons 
via the $^{14}$N(n,p)$^{14}$C reaction 
\citep{goriely00,lugaro03}. 
The $\simeq$~0.001~\msun\ value has been shown in previous studies 
to reproduce observational constraints 
\citep{busso01,cristallo09a}.
Since, in any case, it is a free parameter we also report in Table~\ref{agbtab}
several test cases where the size of this region is 
varied by a factor of 2.
This has only a small effect on the predicted [La/Eu], [Pb/Eu], 
and [Pb/Fe] ratios.

Our method to include the formation of the $^{13}$C pocket 
is very similar to that employed by \citet{goriely00}
and it is based on the simple assumption 
that the proton profile in the $^{13}$C-rich region must be continuous. 
All the mechanisms proposed to date for the mixing 
produce profiles that satisfy this assumption. 
Once this basic feature is assumed, 
the resulting neutron flux and thus the \spro\ distribution 
are almost unequivocally determined 
(except for the two points discussed below).
This was demonstrated by \citet{goriely00} 
who calculated very similar \spro\ distributions 
when changing the shape of the continuous proton profile. 
As a consequence our results are the same as those of 
\citet{vaneck03}, whose models
are based on those of \citet{goriely00}, and those of 
\citet{cristallo09a}, who instead calculated the mixing of 
protons self-consistently via time-dependent overshoot.

Two effects can still change the resulting distribution: 
a higher $^{12}$C abundance, due to overshoot of the 
convective thermal pulses into the C-O core, 
and shear mixing due to rotation occurring 
after the formation of the $^{13}$C pocket. 
A higher $^{12}$C abundance 
would not affect the minimum \spro\ ratios 
because it would result in a higher abundance of $^{13}$C, 
hence a higher neutron flux \citep{lugaro03}, 
a higher Pb/Eu ratio, and unchanged La/Eu ratios. 
Rotational mixing, on the other hand, would completely inhibit 
the \spro\ by mixing $^{14}$N into the $^{13}$C-rich layers of the pocket. 
$^{14}$N would then capture most of the neutrons via the 
$^{14}$N(n,p)$^{14}$C reaction \citep{herwig03,siess04};
however, rotational shear is likely to be damped 
by the inclusion of magnetic fields \citep{suijs08}.
One could still imagine milder mixing leading to smaller neutron exposures. 
\citet{bisterzo10} explicitly investigated this possibility 
by artificially changing the amount of $^{13}$C in the pocket. 
The minimum La/Eu and Pb/Eu adopted here are still valid 
when considering these models. 
Finally, we mention the possibility of the \spro\
occurring during episodes of proton-ingestion 
in the convective thermal pulses 
\citep{cristallo09b}, and 
the minimum values adopted here also hold in this case.

\begin{figure*}
\epsscale{0.85}
\begin{center}
\plotone{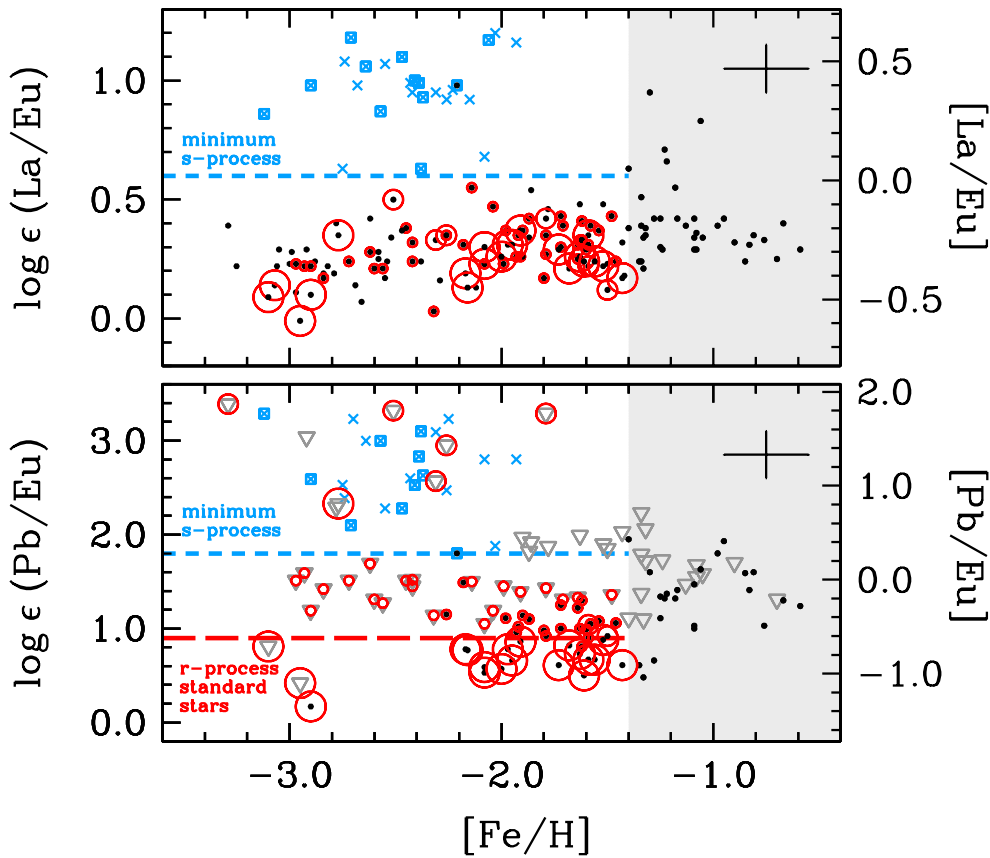}
\end{center}
\caption{
\label{simmererplot}
Logarithmic abundance ratios of La/Eu and Pb/Eu as a function of [Fe/H].
All measurements are indicated by small black circles,
and all upper limits are indicated by downward-facing triangles.
All red circles represent stars lacking any detectable
trace of \spro\ material (see the caption of Figure~\ref{pbeuplot}).
The long-dashed line indicates [Pb/Eu]~$\leq -0.6$ (log~(Pb/Eu)~$\leq +0.9$,
the upper extent of the range for the \rpro\ standard stars),
and the short-dashed lines indicate 
[La/Eu]~$\leq 0.0$
(log~(La/Eu)~$\leq +$0.6) and
[Pb/Eu]~$\leq +0.3$
(log~(Pb/Eu)~$\leq +$1.8),
the approximate minimum ratios expected from AGB pollution.
For comparison, 
small blue ``X''s denote stars enriched in \spro\ material,
and small open squares around these ``X''s indicate that
the star shows RV variations.
The shaded regions indicate metallicities where the
\spro\ predictions may not be appropriate.
A representative uncertainty is shown 
in the top right corner of each panel.
}
\end{figure*}

\subsection{Comparison with Observations}

In addition to displaying stars with \rpro\ enrichment,
Figure~\ref{pbeuplot} also shows
the Pb/Eu ratios for 28 metal-poor stars with reported
\spro\ or $r+s$ enrichments
(\citealt{aoki01,aoki02}, 
\citealt{barbuy05},
\citealt{barklem05},
\citealt{cohen03,cohen06},
\citealt{goswami06},
\citealt{ivans05},
\citealt{johnson04},
\citealt{jonsell06},
\citealt{preston01},
\citealt{roederer08b,roederer10a}, 
\citealt{simmerer04}, and
\citealt{thompson08}).
Members of this group that are in known binary (or multiple) star
systems or have detected radial velocity (RV) variations are 
highlighted 
(see \citealt{aoki03}, \citealt{carney03}, and \citealt{preston09} 
in addition to the above references), 
though the lack of RV variations should
not be taken as strong evidence against binarity \citep{preston09}.\footnote{
The confirmed RV variable stars are preferentially
among those with the highest levels of [Eu/Fe], but this is
to be expected due to observational bias if we assume that
the Eu originated in the \spro.
Stars in close binary systems have shorter periods that 
increase the probability of detecting the RV variations
on shorter timescales.
\citet{boffin94} found a perceptible anti-correlation between 
orbital period and \spro\ enrichment in barium stars
(i.e., Pop~I G--K giants).
In other words, a greater amount of material lost from the
donor star is being captured when the companion is in close proximity,
and this phenomenon is likely manifest here.}
All C-enriched
metal-poor stars with overabundances of \spro\ material
are likely in binary star systems
(e.g., \citealt{mcclure80}, \citealt{mcclure83}, \citealt{lucatello05}).
Most of these stars have 
[Pb/Eu]~$> +$0.3 (log~(Pb/Eu)~$> +$1.8).  
This minimum Pb/Eu ratio is in very good agreement with 
our AGB model predictions.

Figure~\ref{simmererplot} shows both the Pb/Eu and La/Eu
ratios as a function of [Fe/H] for this same sample of 28~stars
with $s$ or $r+s$ enrichment and for our sample of $r$-enriched stars.
The three stars with 0.0~$<$~[La/Eu]~$<$ +0.1 
($+$0.6~$<$~log~(La/Eu)~$< +0.7$)
(\mbox{CS~29513--032}, \mbox{CS~29526--110}, and \mbox{HE~0058$-$0244};
\citealt{roederer10a}, \citealt{aoki02}, and \citealt{cohen06}, respectively)
all have high Pb/Eu ratios ([Pb/Eu]~$> +$1.1), 
so they would not be otherwise mistaken as $r$-enriched.
Furthermore, one of these stars, \mbox{CS~29513--032}, is a member of a
stellar stream with known $r$-enhancement in other stars
(by definition, then, it is an $r+s$ star), so it is 
not surprising that its La/Eu and Pb/Eu ratios have been lowered by the 
presence of \rpro\ material.
On the basis of the RV variability and high La/Eu and Pb/Eu ratios,
it is clear that these stars formed through a separate
enrichment mechanism than the stars that we claim lack
any detectable signature of \spro\ enrichment.
Based on our AGB \spro\ model predictions and the observational
data shown in Figure~\ref{simmererplot}, 
we conservatively adopt 
[La/Eu]~$= 0.0$ (log~(La/Eu)~$= +$0.6) and 
[Pb/Eu]~$= +0.3$ (log~(Pb/Eu)~$= +$1.8)
as the minimum \spro\ ratios expected at low metallicity.

A more concerning scenario is that 
\spro\ material produced by AGB stars has added a light ``dusting''
to the ISM.
Our minimum \spro\ Pb/Eu ratio would need to be diluted by a factor
of $\gtrsim$~10 by the low Pb/Eu ratio found in the \rpro\ standard
stars in order to be disguised as \rpro\ material and
remain undetected by us.
The overwhelming majority of the 28 stars in our 
$s$ and $r+s$ subset have strong C-enhancements
([C/Fe]~$> +$1.5), presumably produced together with the \spro.
Many of the stars that we claim to lack \spro\ material
(``no-$s$'') have subsolar [C/Fe] ratios
(see original source references for Table~\ref{abundtab}, 
especially \citealt{simmerer04}).
To dilute [C/Fe]~$= +$1.5 to a solar [C/Fe] ratio by mixing it
with [C/Fe]~$= -$0.2 (the median value for the sample of stars
analyzed by \citealt{simmerer04}) would 
require a dilution factor of more than 80.
If such dilution is not seen in our sample of no-$s$ stars in [C/Fe], 
it is not likely present in [Pb/Eu] or [La/Eu]. 

\begin{figure*}
\epsscale{0.85}
\begin{center}
\plotone{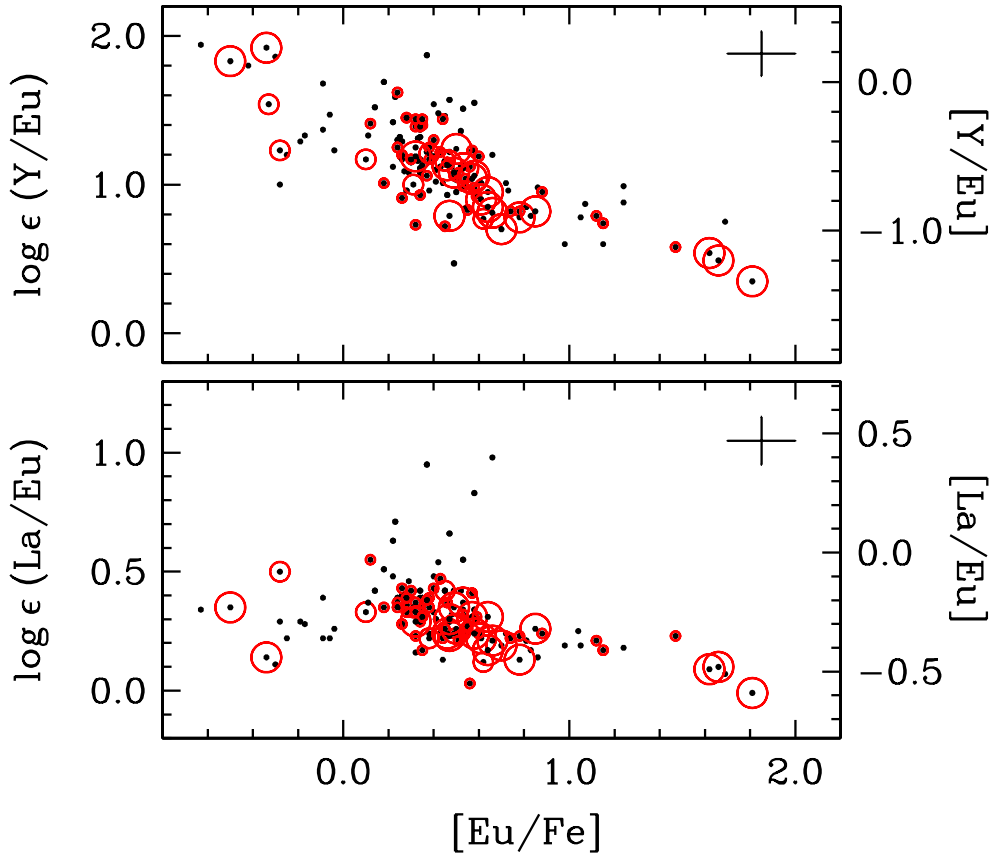}
\end{center}
\caption{
\label{resultplot1}
Logarithmic abundance ratios of 
Y/Eu and La/Eu as a function of [Eu/Fe].
All measurements are indicated by small, filled black circles.
All red circles represent stars lacking any detectable
trace of \spro\ material (see the caption of Figure~\ref{pbeuplot}).
A representative uncertainty is shown 
in the top right corner of each panel.
}
\end{figure*}

Another possible source of an \spro\ dusting 
could be the weak \spro.
This operates in massive stars but is not expected to produce
significant amounts of nuclei heavier than 
$A \simeq$~90 (i.e., the Zr isotopes) \citep{raiteri93},
so this process cannot be the origin of a dusting 
of heavy \ncap\ material.
(This does not exclude the possibility that the weak \spro\ 
may produce some of the $A \lesssim$~90 nuclei 
ejected from a core-collapse SN.)
\citet{pignatari08} present nucleosynthesis calculations
for the weak \spro\ in rotating, massive, low-metallicity 
stars, and their models predict the production of 
heavier \spro\ nuclei; however, even here, the Pb overabundance
is expected to be large.  
In summary, for all but the lightest nuclei,
it seems unlikely that 
\spro\ nucleosynthesis is contributing 
small (or large) amounts of material to our sample of $r$-only stars.

\section{Observed Correlations}
\label{correlations}

For the remainder of this study, we accept
(1) that the small number of well-studied, 
low-metallicity \rpro\ standard stars 
(such as \mbox{CS~22892--052} and \mbox{CS~31082--001})
lack \spro\ material;
(2) that the two low-metallicity stars \mbox{HD~88609} and \mbox{HD~122563},
which are deficient in the heavy \ncap\ elements,
lack \spro\ material; and 
(3) the general presence of high Pb/Fe and Pb/Eu ratios produced 
in the \spro\ at low metallicity.
We now present the resulting
observed heavy element abundance correlations for the $r$-only stars
and discuss their consequences.

\begin{figure*}
\epsscale{0.75}
\begin{center}
\plotone{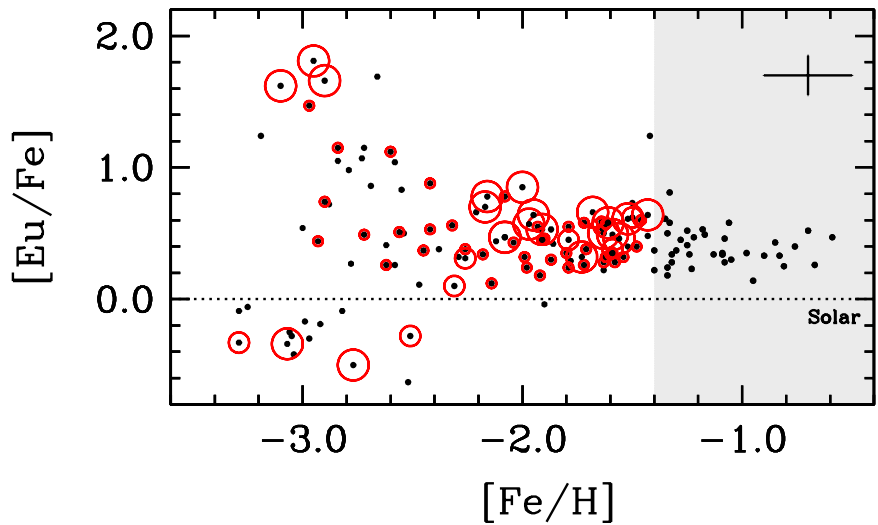}
\end{center}
\caption{
\label{eufeplot}
The [Eu/Fe] ratio as a function of [Fe/H].
Detections are indicated by the small filled circles.
All red circles represent stars lacking any detectable
trace of \spro\ material (see the caption of Figure~\ref{pbeuplot}).
The dotted line indicates the S.S.\ ratio.
The shaded region indicates metallicities where the
\spro\ predictions may not be appropriate.
A representative uncertainty is shown 
in the top right corner.
}
\end{figure*}

In Figure~\ref{simmererplot} we show the logarithmic La/Eu and Pb/Eu 
ratios as a function of [Fe/H] for all stars listed in Table~\ref{abundtab}.
Stars that we have identified as lacking any detectable \spro\ material
are highlighted by the red circles, which we focus on now.
The top panel of Figure~\ref{simmererplot} is analogous to Figure~7 of
\citet{simmerer04}.  
A slight overall upward trend in La/Eu with increasing [Fe/H] 
is apparent, but this is driven by a small number of stars
with low La/Eu near [Fe/H]~$= -$3.0.
\citet{simmerer04} attributed this gradual increase in La/Eu to 
a rise in the amount of \spro\ material present in the birth clouds,
since the high mass stars presumably associated with the \rpro\ 
should have enriched the ISM faster than the lower mass stars
associated with the \spro.
(\citealt{simmerer04} adopted log~(La/Eu)$_{r} \approx +$0.1 and 
log~(La/Eu)$_{s} \approx +$2.1.)
A similar effect is seen for Pb/Eu in the bottom panel of
Figure~\ref{simmererplot}.
Any slope in Pb/Eu is only driven by two stars
with low Pb/Eu at [Fe/H]~$= -$2.9, 
\mbox{CS~31082--001} and \mbox{HE~1523$-$0901};
\mbox{CS~31082--001} is the lone $r$-only star with [Fe/H]~$< -$2.3
and detected Pb.
In the metallicity range from $-$2.3~$<$~[Fe/H]~$< -$1.4, 
there does not appear to be any upward slope in Pb/Eu,
and there is no upward slope in La/Eu in this metallicity range, either.
(Recall that we have refrained from making any assumptions regarding
the origin of the heavy elements in stars with [Fe/H]~$> -$1.4.)

The logarithmic Y/Eu ratio is shown as a function of [Eu/Fe] 
in the top panel of Figure~\ref{resultplot1}.
In the stars lacking any \spro\ material,
there is a marked anti-correlation between [Y/Eu] and [Eu/Fe],
in the sense that the stars with the highest [Eu/Fe] ratios have the
lowest [Y/Eu] ratios.
This anti-correlation is continuous and 
extends several orders of magnitude from
$-$0.5~$\leq$~[Eu/Fe]~$\leq +$1.8 (a factor of 200 in Eu/Fe)
and includes the stars most strongly enriched in the \rpro\ 
(e.g., \mbox{CS~22892--052}) and those with the most
severe heavy element deficiencies (e.g., \mbox{HD~122563}).
There is a fair amount of scatter in the relation
(a factor of $\sim$~2--8 in Y/Eu), increasing
in Y/Eu with decreasing [Eu/Fe],
but this scatter is much smaller than the extent over which the
relationship extends (a factor of $>$~30 in Y/Eu).
The existence of this relationship
reaffirms the findings of \citet{barklem05}, \citet{otsuki06},
and \citet{montes07}
on the basis of a more extensive set of stellar abundances
that has been explicitly purged of \spro\ contamination.

The bottom panel of Figure~\ref{resultplot1} shows the logarithmic La/Eu 
ratio as a function of [Eu/Fe].
There is a hint of an anti-correlation between these variables---but 
it is not nearly as pronounced as the relationship between 
[Y/Eu] and [Eu/Fe].
This relationship helps to explain the slight
upward trend of La/Eu with increasing [Fe/H] 
seen in the top panel of Figure~\ref{simmererplot}.
The stars with the lowest La/Eu ratio are generally those with
the highest levels of [Eu/Fe], which preferentially occur
in stars with [Fe/H]~$< -$2.5, as shown in Figure~\ref{eufeplot}.
At higher metallicities [Eu/Fe] is generally lower and thus
La/Eu is slightly higher, so the upward trend of 
La/Eu with increasing [Fe/H] in Figure~\ref{simmererplot} is
not explicitly a metallicity effect.\footnote{
Figure~\ref{resultplot1} also reveals several stars with super-Solar
[La/Eu] ratios, including 3~stars with [La/Eu]~$> +0.2$:
\mbox{BD~$-$01~2582}, \mbox{G126-036}, and \mbox{G140-046}.
All of these stars except \mbox{BD~$-$01~2582} have 
[Fe/H]~$> -$1.4, so they fall beyond
the realm of concern for this study.
\mbox{BD~$-$01~2582}, with [Fe/H]~$= -$2.2, 
is a well-known CH giant \citep{bond80}, and
\citet{carney03} demonstrated that this star exhibits RV variations.
On the basis of its large C enhancement and RV variations
\mbox{BD~$-$01~2582} would not be mistaken for an
$r$-only star, but our derived Pb/Eu ratio for this star,
log~(Pb/Eu)~$= +$1.80, places it squarely on our adopted lower
limit for AGB pollution.
This star serves as a cautionary reminder to consider 
all available evidence when examining the enrichment history of a star.}

\begin{figure*}
\epsscale{0.85}
\begin{center}
\plotone{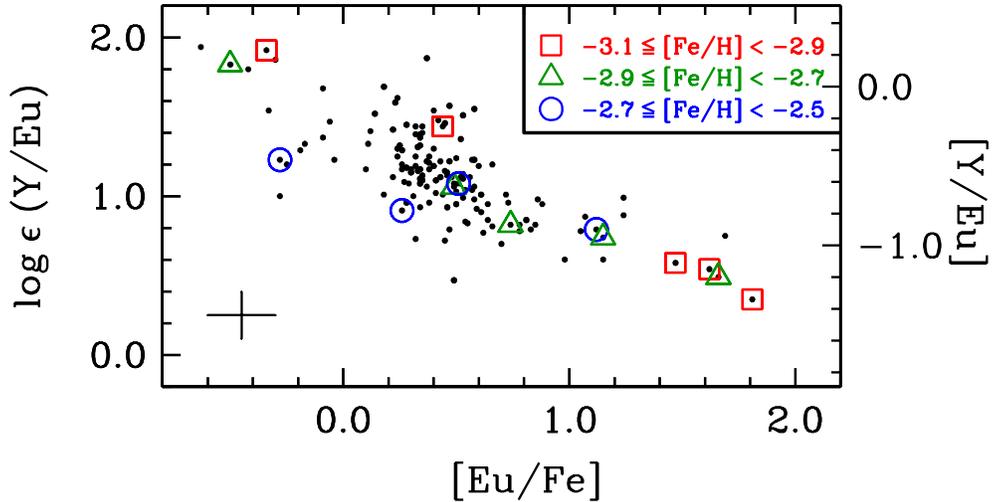}
\end{center}
\caption{
\label{resultplot3}
The logarithmic Y/Eu ratio as a function of [Eu/Fe], 
which is the same as in Figure~\ref{resultplot1}.
All measurements are indicated by small black circles.
Stars with no detectable trace of \spro\ material
that have metallicities between $-3.1 \leq$~[Fe/H]~$< -2.5$
are highlighted as indicated in the figure key.
Representative uncertainties are shown 
in the lower left corner.
}
\end{figure*}

Figure~\ref{resultplot3} illustrates this fact explicitly.
In each of three metallicity bins
([Fe/H]~$= -$3.0~$\pm$~0.1, 
 [Fe/H]~$= -$2.8~$\pm$~0.1, and
 [Fe/H]~$= -$2.6~$\pm$~0.1)
there are several stars whose [Eu/Fe] ratios span most or all of
the observed range
($-$0.4~$<$~[Eu/Fe]~$< +$1.8, 
$-$0.5~$<$~[Eu/Fe]~$< +$1.6, and
$-$0.3~$<$~[Eu/Fe]~$< +$1.1, respectively).
This firmly indicates that the relationship between 
[Eu/Fe] and [Y/Eu] is independent of metallicity,
which also reaffirms the findings of \citet{montes07}
(their Figure~2).

\begin{figure*}
\epsscale{0.85}
\begin{center}
\plotone{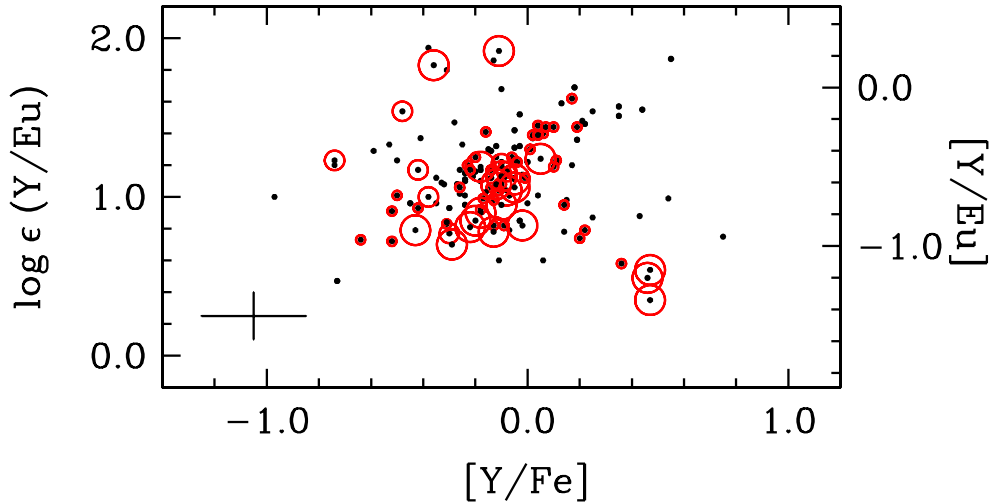}
\end{center}
\caption{
\label{resultplot2}
The Y/Eu ratio as a function of [Y/Fe].
All red circles represent stars lacking any detectable
trace of \spro\ material (see the caption of Figure~\ref{pbeuplot}).
A representative uncertainty is shown 
in the lower left corner.
}
\end{figure*}

Figure~\ref{resultplot2} compares the [Y/Eu] ratio to [Y/Fe].
The [Y/Fe] ratio is super-Solar in the handful of stars
with [Eu/Fe]~$> +$1.0, but
in all other cases there appears to be no relationship between
[Y/Eu] and [Y/Fe].
Unlike the top panel of Figure~\ref{resultplot1}, where
[Y/Eu] showed a clear anti-correlation with [Eu/Fe]
spanning the entire range of [Eu/Fe],
there is no relationship between [Y/Eu] and [Y/Fe] except
for the most $r$-rich stars.
When Eu is produced in significant quantities
([Eu/Fe]~$> +$1.5), Y is also produced
in slightly higher amounts as well ([Y/Fe]~$> +$0.4).
On the other hand, when lower amounts of Y are produced 
([Y/Fe]~$<$~0), the amount of Eu produced may vary 
by more than 1~dex for a given abundance of Y.
Knowing [Y/Fe] for a star gives little predictive power 
for the [Y/Eu] ratio, whereas [Eu/Fe] does.
Stars strongly enriched by the \rpro, such as \mbox{CS~22892--052},
are overabundant in the heavy elements relative to the light ones, and 
stars such as \mbox{HD~122563} are deficient in the heavy elements,
rather than overabundant in the light ones.

\begin{figure*}
\epsscale{0.7}
\begin{center}
\plotone{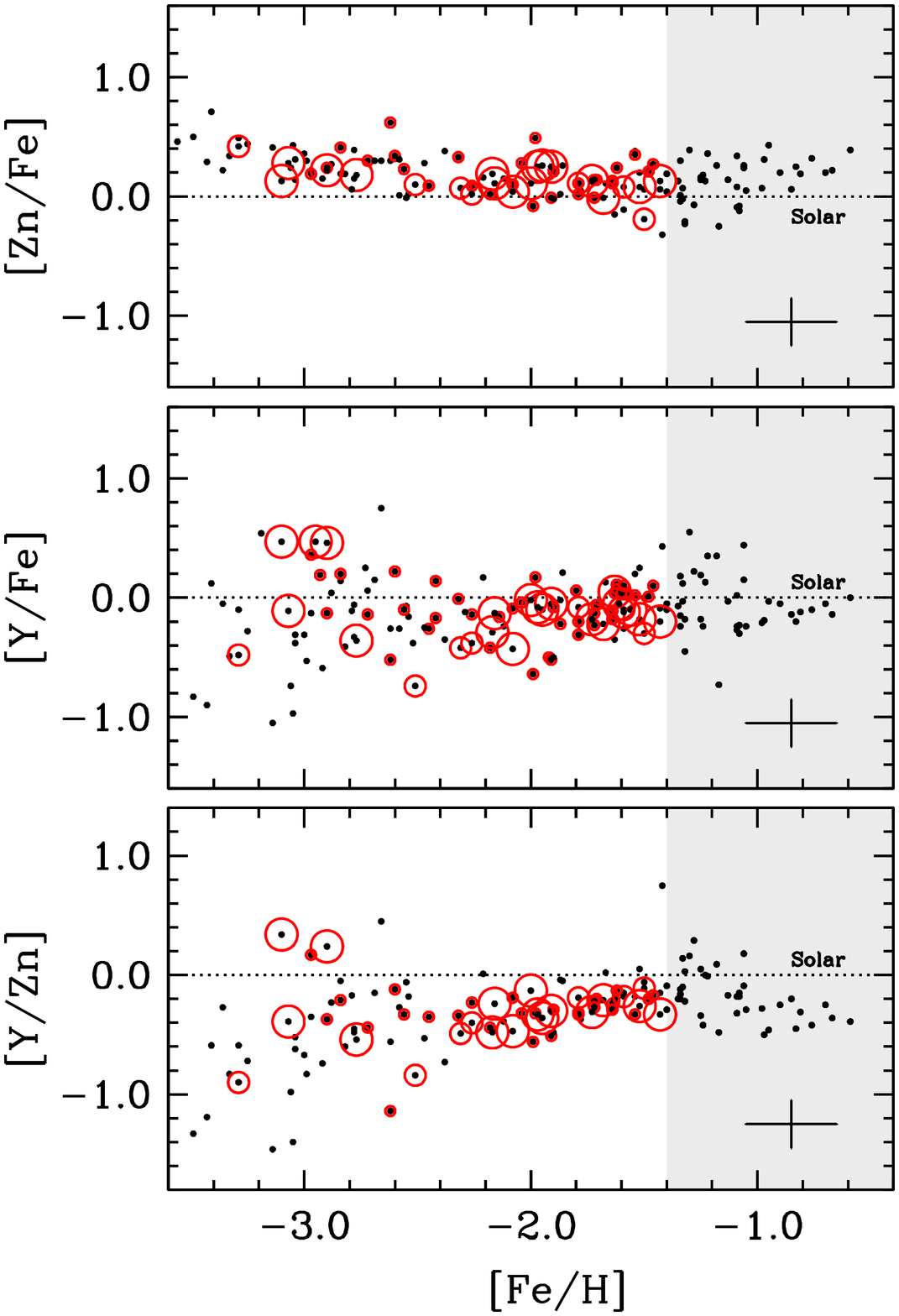}
\end{center}
\caption{
\label{znyplot}
The [Zn/Fe], [Y/Fe], and [Y/Zn] ratios as a function of [Fe/H].
All red circles represent stars lacking any detectable
trace of \spro\ material (see the caption of Figure~\ref{pbeuplot}).
Dotted lines indicate the S.S.\ ratios.
The shaded regions indicate metallicities where the
\spro\ predictions may not be appropriate.
A representative uncertainty is shown 
in the lower right corner of each panel.
}
\end{figure*}

At low metallicity, elements at least as heavy as Ge ($Z =$~32)
are produced along with the Fe-group and not in \ncap\ reactions
(\citealt{cowan05}; see also \citealt{frohlich06} and \citealt{farouqi09}).
Figure~\ref{znyplot} demonstrates that the Y in our sample
is clearly decoupled from the Fe-group elements Fe and Zn.
The [Zn/Fe] ratio shows almost no scatter at all metallicities
in these stars and has a slight upturn at [Fe/H]~$< -$2.8.
Zn, the heaviest element in the Fe-group that is easily 
measured in metal-poor stars, is clearly produced along with Fe.
In contrast, the [Y/Fe] ratio shows an increasingly large degree
of scatter at low metallicities.
Knowing the Zn (or Fe) abundance of a star gives no indication
of the Y abundance and vice-versa, indicating that the Y in our
sample was not produced with the Fe-group elements.

Having shown that we can select a sample of stars with no \spro\ 
enhancement and having identified a relationship between the
abundance ratios in these stars, we now 
propose a mechanism to explain this relationship.

\section{Heavy Element Nucleosynthesis in the High Entropy Wind
of a Core-Collapse SN}
\label{hew}

Despite many years of effort, the specific astrophysical site for the
\rpro\ is still unknown;
core-collapse supernovae (SNe), however, have long been suspected
as one promising source for this process,
despite the difficulty in understanding (and replicating) the
explosion mechanism and exotic SN physics.
Model-independent approaches have been 
utilized to attempt to characterize the nature of the \rpro\
in explosive environments. 
These ``waiting point'' approximation models, 
based on the neutron capture and photodisintegration equilibrium 
under conditions of high neutron number densities, 
have provided insight into the 
nuclear and astrophysical conditions necessary for the \rpro\
(see \citealt{kratz93,kratz07}).
To synthesize neutron-rich nuclei in explosive environments
requires some combination of values of 
neutron number densities or entropies (S). 
One promising SN model involves the so-called ``neutrino wind,''
a wind of particles caused by neutrinos
shortly after the SN explosion
(see, e.g., \citealt{woosley94,thompson03}).
This scenario posits a moderately neutron-rich,   
high entropy wind (HEW) from Type~II (core-collapse) SNe
(see also \citealt{wanajo02}).
To explore the nucleosynthetic conditions in this HEW,
\citet{farouqi09,farouqi10} have performed a number of nucleosynthesis 
network calculations to determine the ratio of
free neutrons to ``seed'' nuclei (Y$_n$/Y$_{\rm seed}$), 
which is correlated with entropy, the electron abundance 
Y$_e =$~($Z/A$), and the expansion velocity. 
Hydrodynamical simulations cannot yet
reproduce the detailed astrophysical and 
nuclear conditions in the SN explosion,
but it is possible to explore the parameter space  
in our HEW simulations with different values of S and Y$_e$
to determine the ratio Y$_n$/Y$_{\rm seed}$,
which can be thought of as the strength of the \rpro\ 
\citep{kratz08b,farouqi09,farouqi10}.

The term ``\rpro'' may describe one particular nucleosynthetic 
mechanism for producing heavy nuclei (specifically, the 
addition of large numbers of neutrons to existing nuclei
on timescales much shorter than the $\beta$-decay rates), but 
the conditions that enable such a process may 
span a wide range of physical properties that
together may be capable of producing a range of abundance patterns.
This is revealed in the results of both the waiting point approximation 
and the HEW model calculations. 
\citet{kratz07}, for example, find that different neutron number densities 
are required to produce different abundance regimes. 
They could reproduce the S.S.\ \rpro\ abundance curve
and the $r$-rich halo star elemental abundances with a superposition of 
neutron number densities ranging from 
20~$\leq$~log n$_n \leq$~28. 
The heavier \ncap\ elements ($A \gtrsim$~130, roughly the Ba isotopes
and heavier)
required 23~$\leq$~log n$_n \leq$~28, typical of
the main \rpro, while the lighter elements could be reproduced
with only 20~$\leq$~log n$_n \leq$~22.
In more sophisticated HEW dynamic network calculations,
\citet{farouqi09} found that a superposition of weighted
entropies for a fixed Y$_e =$~0.45
was necessary to reproduce the S.S.\ \rpro\ abundance curve
and $r$-rich halo stars:
the $A \gtrsim$~130 nuclei could be produced with 150~$<$~S~$<$~300
(typical of the main \rpro), but the $A \lesssim$~130 nuclei
required only 110~$<$~S~$<$~150 (typical of the weak 
component of the \rpro\ as defined by \citealt{pfeiffer01,truran02}, 
which does not produce the Ba isotopes).

Here we compare observations with 
recent dynamic \rpro\ simulations in the HEW,
assuming the full entropy range
(5~$\leq$~S~$\leq$~300, which depends on Y$_e$; see \citealt{farouqi10}) 
and an expansion velocity of 7500~\kmsec.
These new calculations employ the Extended Thomas Fermi 
mass model with quenched shell effects (ETFSI-Q) far 
from stability to predict masses where no experimental data are
available.  
Furthermore,  
the nuclear physics input parameters,   
including the half lives, \ncap\ cross sections,
$\beta$-delayed neutron emission probability,
and fission rates have all been obtained {\it consistently} based upon the 
same ETFSI-Q model (see \citealt{farouqi10} for further discussion).  

\begin{figure*}
\epsscale{0.85}
\begin{center}
\plotone{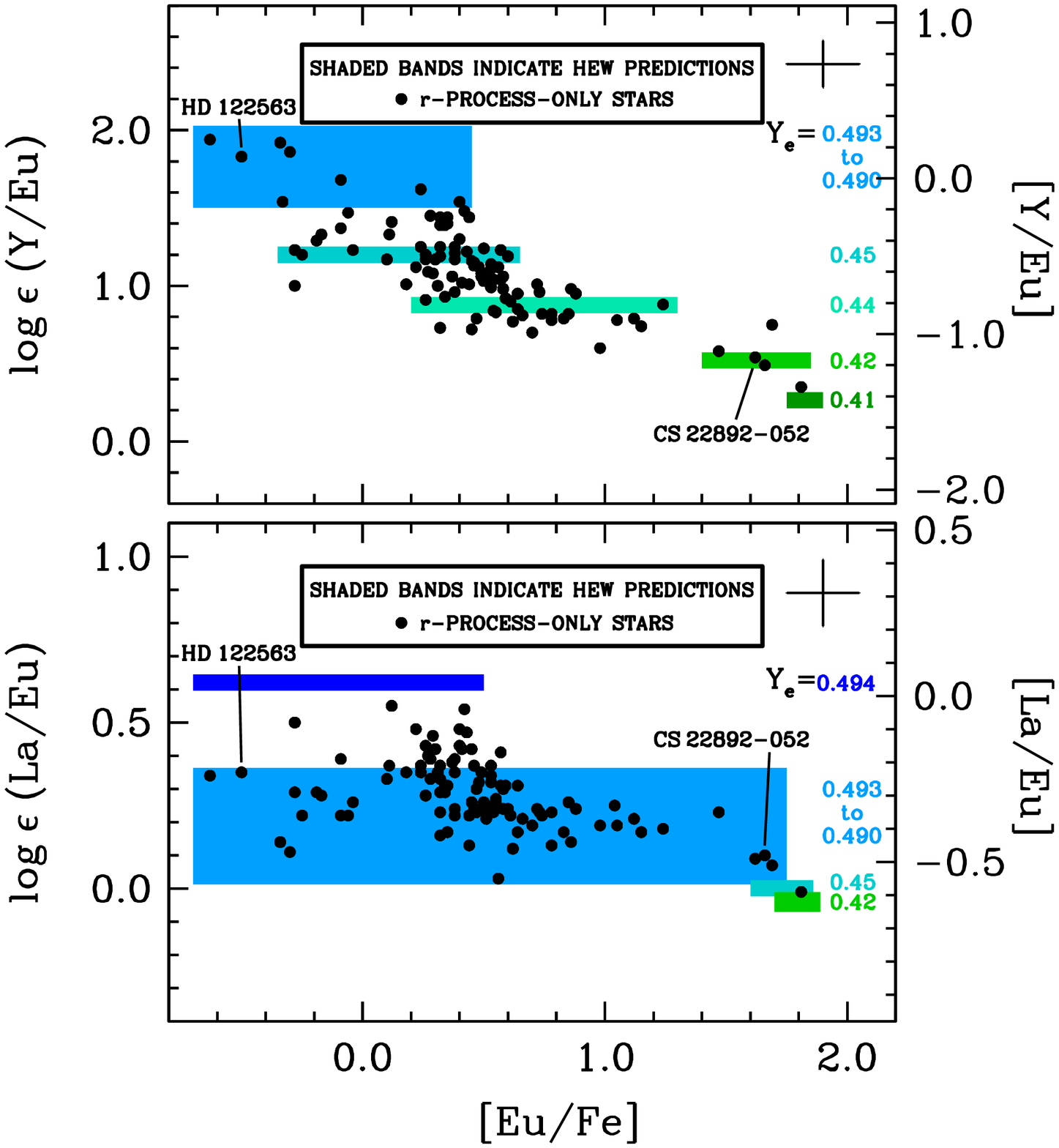}
\end{center}
\caption{
\label{hewplot}
Logarithmic Y/Eu and La/Eu ratios as a function of [Eu/Fe] for 
stars with [Fe/H]~$< -$1.4 and 
[La/Eu]~$< 0.0$ (log~(La/Eu)~$< +$0.6; see Figure~\ref{simmererplot}).
The shaded bands indicate different Y$_e$ ranges from our HEW simulations
assuming the full entropy range (see Table~4 of \citealt{farouqi10}) 
and an expansion velocity of 7500~\kmsec.
The HEW simulations do not predict [Eu/Fe] explicitly, so the horizontal
ranges are scaled to approximately match the observational data.
A representative observational uncertainty is shown 
in the upper right corner of each panel.
}
\end{figure*}

Figure~\ref{hewplot} shows the logarithmic Y/Eu and La/Eu ratios
as a function of [Eu/Fe] for all stars in our sample with
[La/Eu]~$<$~0.0 (log~(La/Eu)~$< +$0.6).
As shown in the top panel, the Y/Eu ratio in the
\rpro\ rich star \mbox{CS~22892--052}
is 25~times smaller than that in the \rpro\ deficient star
\mbox{HD~122563}.
These extreme ratios can be matched simply by varying Y$_e$
in our calculations from $\approx$~0.49 for \mbox{HD~122563}
to 0.42 for \mbox{CS~22892--052}.
The \textit{vertical} placement of the Y$_e$ values 
(i.e., [Y/Eu] or log~(Y/Eu)) in Figure~\ref{hewplot} is explicitly predicted
by our simulations.
We caution that the \textit{horizontal} placement (i.e., [Eu/Fe])
of the Y$_e$ bands is not an explicit prediction, and the 
horizontal extent of the bars has been scaled to 
approximately match the observational data.

\begin{figure*}
\epsscale{0.70}
\begin{center}
\plotone{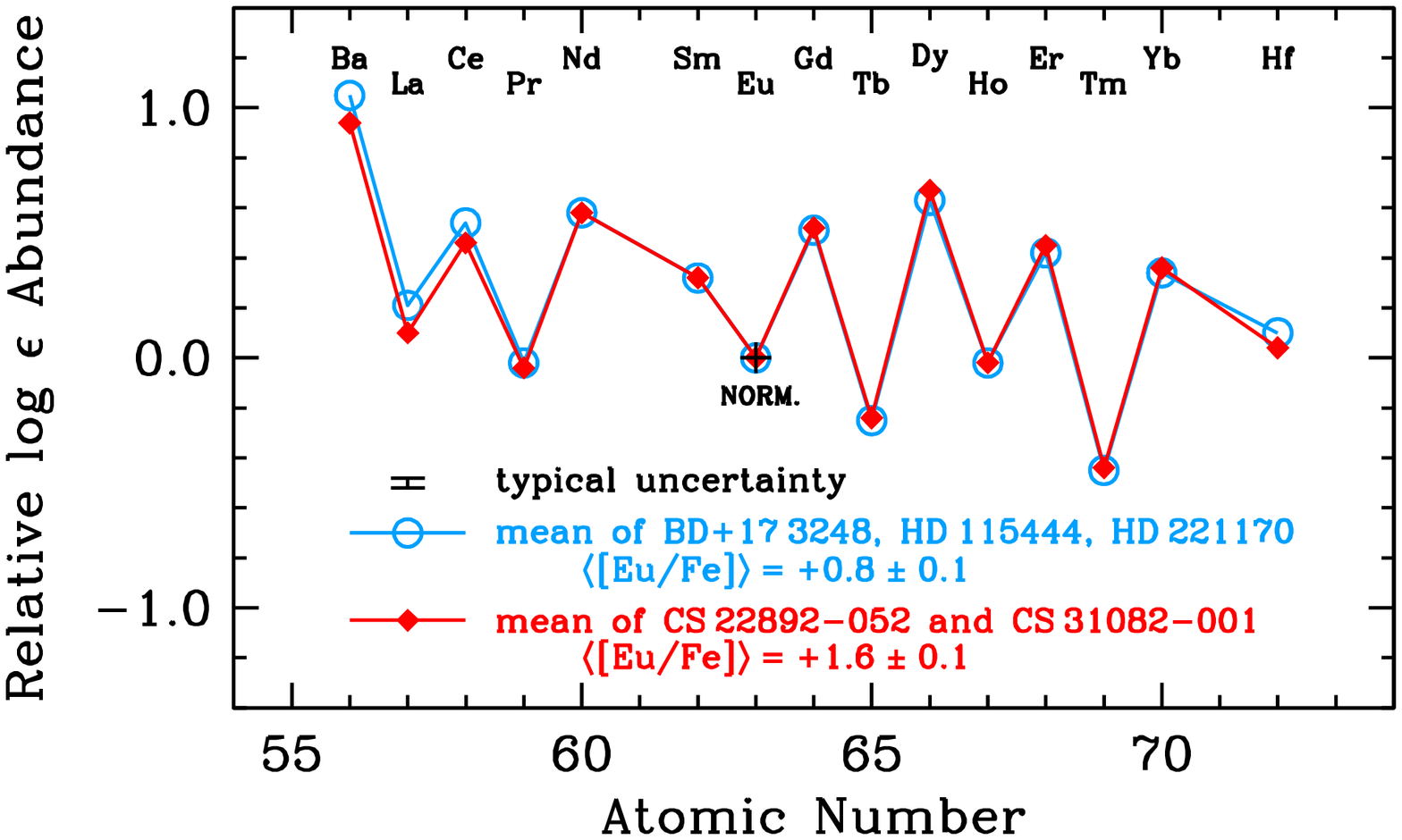}
\end{center}
\caption{
\label{sneden09plot}
Relative mean abundances for the REE
in two groups of \rpro\ standard stars.
The first group (blue circles) has $\langle$[Eu/Fe]$\rangle = +$0.8~$\pm$0.1:
\mbox{BD$+$17~3248}, \mbox{HD~115444}, and \mbox{HD~221170}.
The second group (red diamonds) has $\langle$[Eu/Fe]$\rangle = +$1.6~$\pm$0.1:
\mbox{CS~22892--052} and \mbox{CS~31082--001}.
The abundances are normalized to Eu ($Z =$~63).
Abundances are taken from \citet{sneden09}.
There is a notable difference in the abundances of the three
lightest REE (Ba, La, and Ce).
}
\end{figure*}

In the lower panel of Figure~\ref{hewplot}, 
the logarithmic La/Eu ratio shows a relatively flat trend with
a comparatively small change in Y$_e$ from the 
$r$-deficient to the $r$-rich stars,
with Y$_e >$~0.49 for \mbox{HD~122563} to
$\approx$~0.49 for \mbox{CS~22892--052}.
For 0.40~$<$~Y$_e <$~0.49, the [La/Eu] predicted
by our simulations changes only by $<$~0.1~dex
(in contrast to a change of [Y/Eu] of $\approx$~1.8~dex).
When using the ETFSI-Q mass model, our HEW predictions for
the light REE are $\sim$~0.2~dex too low compared with,
e.g., the \citet{arlandini99}
S.S.\ $r$-residuals.
This results from the well-understood nuclear structure deficiencies
in the transitional region beyond the $N =$~82 shell closure, which
affect the neutron separation energies and consequently the
\rpro\ path.
Most other mass models show even more significant deficiencies
than the ETFSI-Q model.
If we ``repair'' the ETFSI-Q model in this
region (i.e., artificially add the 0.2~dex), we recover the
same Y$_e$ fractions for both La/Eu and Y/Eu:
the lowest stellar [La/Eu] ratios at $\approx -$0.6 would be reproduced
with Y$_e =$~0.41, the highest stellar [La/Eu] ratios at 
$\approx -$0.05 would be reproduced by Y$_e =$~0.493, and all intermediate
ratios would be shifted up accordingly.
Alternatively, these observed ranges in 
[Y/Eu] and [La/Eu] may also be fit by fixing Y$_e =$~0.45 
and varying the entropy ranges
(e.g., from 5~$\leq$~S~$\leq$~215 for the $r$-deficient stars
to 70~$\leq$~S~$\leq$~300 for the $r$-rich stars).\footnote{
Increasing the entropy range from S~$\leq$~230
to S~$\leq$~300 changes La/Eu 
very little, and removing the low entropy components from the HEW
calculations affects the abundances of each of La and Eu by $\ll$~1\%
even when S~$\leq$~175 are removed.
In other words, neither La nor Eu are being produced in significant
quantities until the \rpro\ flow has passed the closed nuclear shells
that produce the $A \sim$~130 abundance peak.
The REE, including La and Eu, are 
produced under similar Y$_n$/Y$_{\rm seed}$ conditions
within a small entropy interval.}

Thus our HEW simulations 
can successfully reproduce both the Y/Eu and La/Eu ratios
for both the $r$-rich and $r$-deficient stars 
(as well as the intermediate cases)
with self-consistent ranges of Y$_e$ or entropy.
A robust main \rpro\ produces abundance patterns like those
seen in \mbox{CS~22892--052} with low Y/Eu ratios.
Stars like \mbox{HD~122563}, with a higher Y/Eu ratio matched by
a higher Y$_e$ (e.g., Figure~2 of \citealt{kratz08a}),
can be considered to be enriched by an incomplete main \rpro\ 
where the production
of the heavier elements is falling off with increasing atomic number.

The simulations and abundance comparisons 
do provide some indications of the types of environments 
where this nucleosynthesis may have occurred.
The neutrino-driven wind starts from the surface of the 
proto-neutron star with a flux of neutrons and protons.
As the nucleons cool they combine to form $\alpha$ particles and
an excess of unbound neutrons, and further cooling produces
a population of Fe-group seed nuclei
(e.g., \citealt{woosley94}, \citealt{woosley05}, \citealt{farouqi10}).
For $S \leq$~110 (at fixed Y$_e =$~0.45), 
where the ratio of free neutrons to seed nuclei is $<$~1,
the nucleosynthesis is consistent with a charged-particle (CP)
or $\alpha$-rich freezeout and recapture of $\beta$-delayed neutrons
emitted from neutron-rich nuclei near the first \rpro\ peak.
In this sense, these low entropy components that produce
the Sr-Y-Zr group are of a primary 
nature and fit the requirements for the light element primary
process (LEPP) proposed by \citet{travaglio04}.
(See \citealt{kratz08b} and \citealt{farouqi09}, 
who showed that the Sr/Y/Zr ratios---both
observationally and in the HEW model---are 
independent of total Eu enrichment.)
Identification of the mass range where the production mechanism
changes from a CP and $\beta$-delayed neutron recapture process
to a true \rpro\ is beyond the scope of the present study.

We stress that the relationship between [Y/Eu] and [Eu/Fe]
in metal-poor stars is an observed trend, and the HEW model is 
one plausible explanation for the existence of such a relationship.
This does not, however, exclude the possibility that additional sites---and 
processes---may also produce conditions favorable to heavy element 
nucleosynthesis.
Regardless of which site(s) is (are) responsible for producing the \rpro, 
nuclear physics and realistic astrophysical conditions will remain
essential ingredients to interpreting observed stellar abundance patterns.

\section{Discussion}
\label{discussion}

In this study we adopt somewhat conservative limits
that \rpro\ nucleosynthesis is characterized by 
[Pb/Eu]~$< +$0.3 (log~(Pb/Eu)~$< +$1.8) and
[La/Eu]~$<$~0.0  (log~(La/Eu)~$< +$0.6), from which a correlation
between [Y/Eu] and [Eu/Fe] has emerged.
We are encouraged by the fact that even weak upper limits
on the Pb abundance can sometimes be meaningful.
In this section we consider several examples of how these definitions can 
be used in conjunction with other information to characterize
the heavy element enrichment patterns in metal-poor stars.
We also consider several implications of these results.

\subsection{The Limits of Precision of $r$-process Residuals}

\begin{figure*}
\begin{center}
\includegraphics[angle=270,width=4.5in]{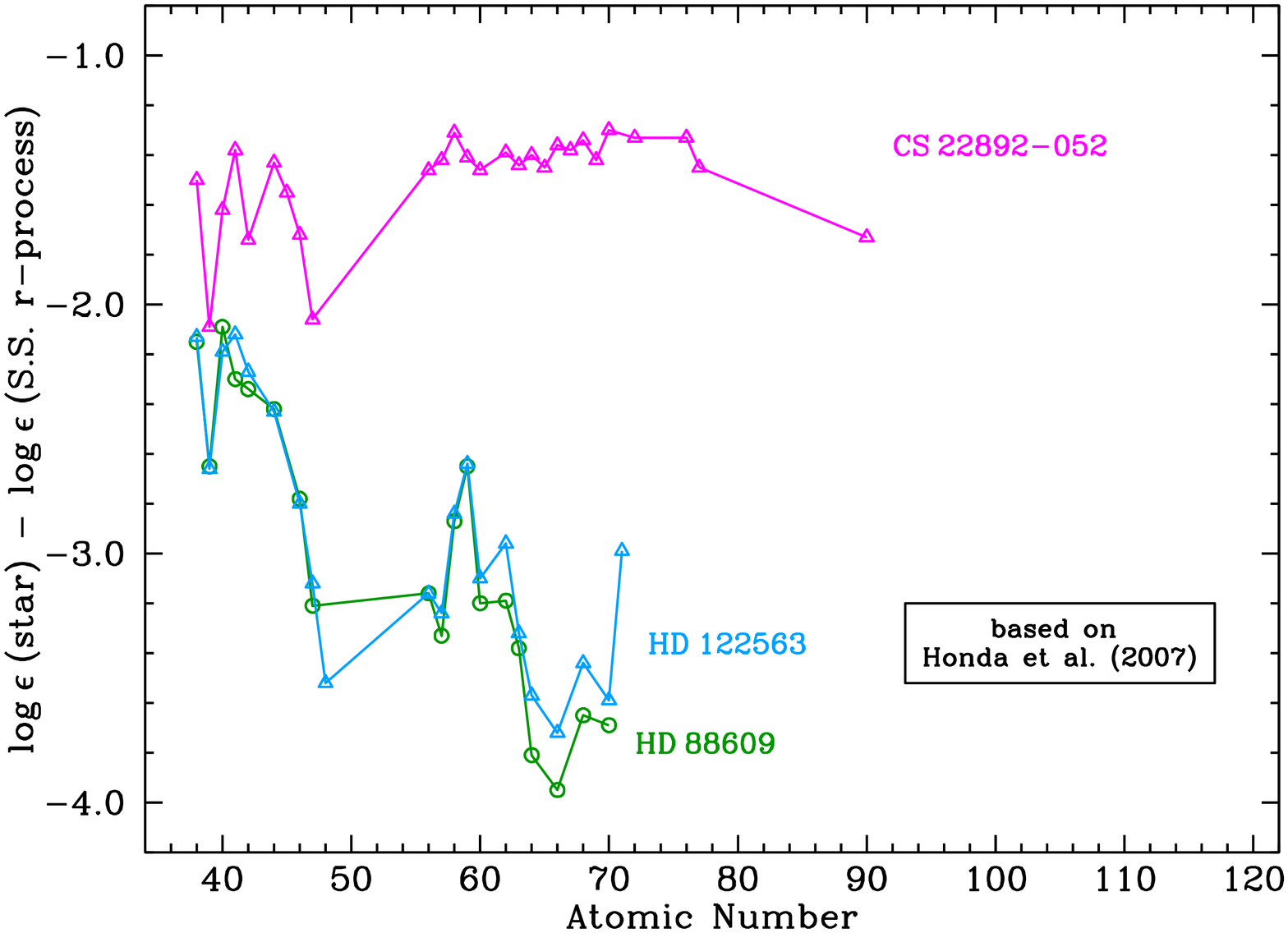} \\
\vspace*{0.3in}
\includegraphics[angle=270,width=4.5in]{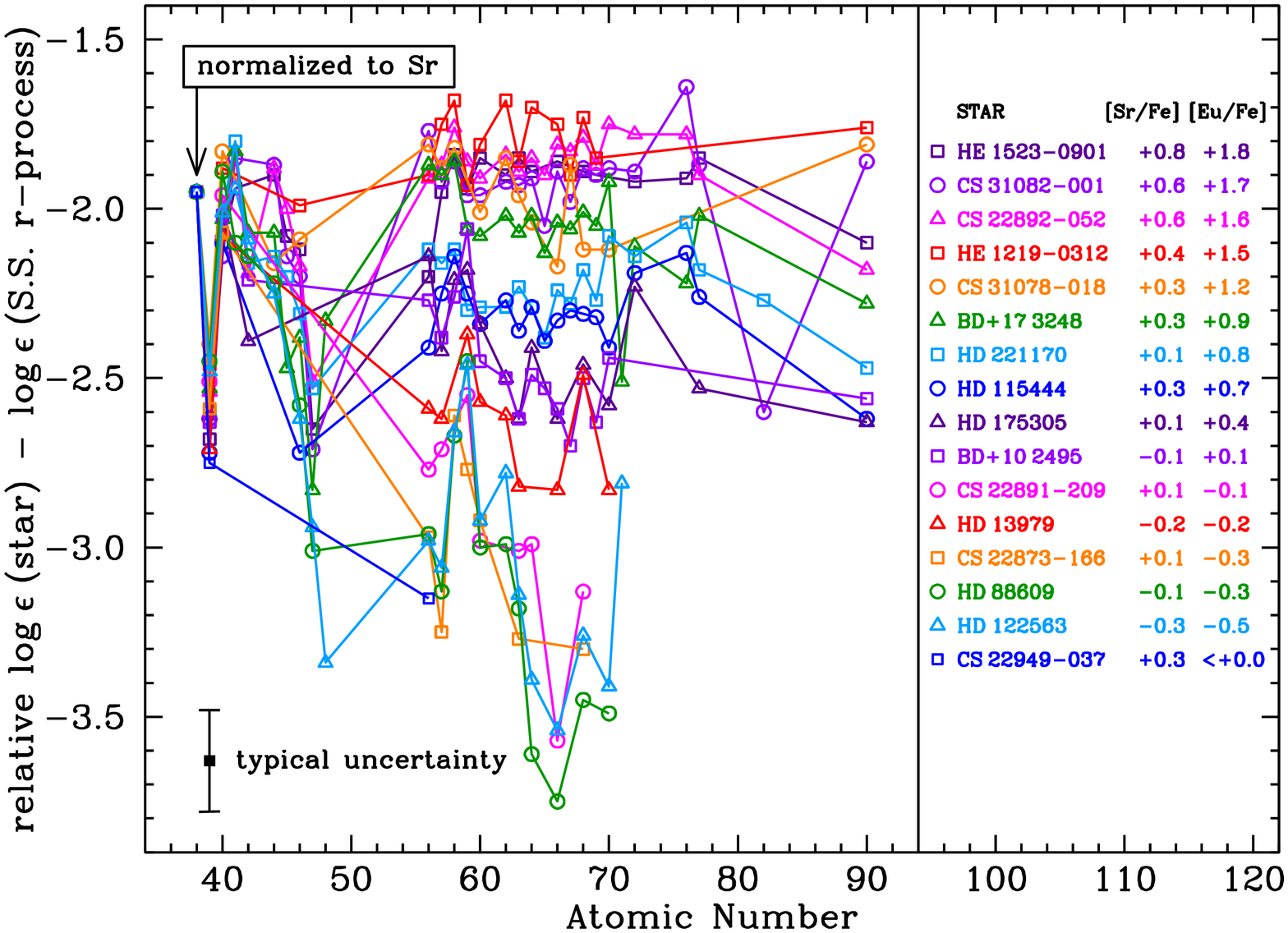}
\end{center}
\clearpage
\caption{
\label{rprostars1}
\scriptsize
\textit{Top panel:} 
Differences between the S.S.\ \rpro\ abundances and
stellar abundances for 3~metal-poor stars, based on Figure~5 
of \citet{honda07}.
Note the clear separation between the two groups of stars.
\textit{Bottom panel:}
Differences between the S.S.\ \rpro\ abundances and
stellar abundances for 16~metal-poor stars, normalized to 
Sr ($Z =$~38) to minimize the effect of overall metallicity differences.
The stars are listed according to decreasing [Eu/Fe] and are
identified, along with their
[Sr/Fe] and [Eu/Fe] ratios, in the box to the right.
A typical uncertainty is indicated in the lower left corner.
This confirms the conclusion of \citet{honda07} that the
heavy element abundance pattern of \mbox{CS~22892--052} is
distinct from \mbox{HD~88609} and \mbox{HD~122563}.
Furthermore, there is a continuous distribution of 
abundance patterns in other stars that fall between these two extremes.
Abundance references are as follows:
S.S.\ \rpro\ abundances, \citet{sneden08};
\mbox{HE~1523--0901}, \citet{frebel07} 
   and A.\ Frebel (2009, private communication);
\mbox{CS~31082--001}, \citet{hill02}, \citet{plez04}, and \citet{sneden09};
\mbox{CS~22892--052}, \citet{sneden03,sneden09};
\mbox{HE~1219--0312}, \citet{hayek09} and \citet{roederer09b};
\mbox{CS~31078--018}, \citet{lai08};
\mbox{BD$+$17~3248}, \citet{cowan02}, \citet{roederer09b}, 
   and \citet{sneden09};
\mbox{HD~221170}, \citet{ivans06} and \citet{sneden09};
\mbox{HD~115444}, \citet{westin00}, \citet{roederer09b}, and \citet{sneden09};
\mbox{HD~175305}, \citet{roederer10a};
\mbox{BD$+$10~2495}, \citet{roederer10a};
\mbox{CS~22891--209}, \citet{francois07};
\mbox{HD~13979}, I.\ Roederer et al., in preparation;
\mbox{CS~22873--166}, \citet{francois07};
\mbox{HD~88609}, \citet{honda07};
\mbox{HD~122563}, \citet{honda06} and \citet{roederer10b};
\mbox{CS22949--037}, \citet{depagne02}.
}
\end{figure*}

Several mechanisms are required to explain the lightest of the
heavy elements in metal-poor stars, and it is now
well-established that simple \rpro\ residuals 
($N_{\rm S.S., r} \equiv N_{\rm S.S., total} - N_{\rm S.S., s}$)
are inadequate descriptions of the 
\rpro\ contribution to the Sr-Y-Zr group
(see \citealt{qian07,qian08} for recent summaries).
Similarly, simple
linear combinations of the scaled S.S.\ \spro\ and \rpro\ are inadequate 
descriptions of some of the \textit{heavy} \ncap\ elements, as well,
when a precise deconvolution is desired.
The observed dispersion in \rpro\ yields must be accounted for.
The \rpro\ dispersion in [La/Eu] is at least 0.5~dex in stars
with [Eu/Fe]~$\lesssim +$0.5,
though the dispersion decreases with increasing [Eu/Fe].
For the $r$-rich stars 
(such as our \rpro\ standards discussed in Section~\ref{intro})
the La/Eu ratio is remarkably constant 
(to a precision of about 0.1~dex; see Section~\ref{examples} below), 
indicating that 
when the \rpro\ fully ``flows'' the heavy elements are produced
in relatively constant ratios.\footnote{
Furthermore, a robust \rpro\ that replicates the third \rpro\ peak
(either in the HEW model or in the waiting point approximation models)
completely produces the actinides, such as Th and U, 
resulting in relatively constant Th/Eu production values.
This reaffirms the reliability of using these element pairs as chronometers.}

The standard method of computing \rpro\
residuals (or ``pure'' $s$- or \rpro\ ratios between two elements)
is of course
still adequate for assessing the relative dominance of the
$s$- or \rpro\ in a general sense, but 
for precision analyses greater caution is warranted.

\subsection{Recognizing \rpro\ Nucleosynthesis in Metal-Poor Stars}
\label{examples}

Small variations in the \rpro\ abundance pattern may be observed
\textit{within} the REE domain.\footnote{Here we expand the REE domain
beyond the lanthanides to encompass Ba through Hf ($Z =$~56--72).} 
For example, in Figure~\ref{sneden09plot} the
REE abundance distribution of 5 \rpro\ 
standard stars are intercompared.  
The lightest REE
(Ba, La, and Ce) in 
\mbox{BD~$+$17~3248}, \mbox{HD~115444}, and \mbox{HD~221170}
have higher mean abundances than
the other two stars, \mbox{CS~22892--052} and \mbox{CS~31082--001}.
These are differences of $\approx$~0.10~dex, while the 
standard deviation of the mean in each group is $\approx$~0.02~dex.
The first group of stars all have [Eu/Fe]~$= +$0.8~$\pm$~0.1
while the second group have [Eu/Fe]~$= +$1.6~$\pm$~0.1.
The stars in the first group  have metallicities 
[Fe/H]~$= -$2.1, $-$2.9, and $-$2.2, respectively, while the
stars in the second group have metallicities
[Fe/H]~$= -$2.9 and $-$3.1, indicating that this is not
explicitly a metallicity effect.
A similar effect can be observed in Figure~13 of \citet{roederer10a},
where Ba--Nd all are slightly overabundant relative to the heavier
REE.
This demonstrates that even in cases where the \rpro\ produces
a large overabundance of heavy material (relative to the 
Fe-group seeds), slight variations can be identified and characterized.

This result affirms that the heavy element abundance pattern in the star 
\mbox{HD~126238} 
can be explained through enrichment by only the \rpro.
This star, reviewed previously in Section~\ref{nos}, 
has a low Pb abundance (log~(Pb/Eu)~$= +$1; \citealt{cowan96}), 
and thus there is no need to invoke an \spro\ dusting of material.

One of the stars in the stellar stream
analyzed by \citet{roederer10a}, \mbox{HD~175305},
was included in the study of \citet{roederer08a} of
the isotopic fractions of three REE (Nd, Sm, and Eu).
The excess Ba and Ce relative to the scaled S.S.\ \rpro\ pattern
was interpreted as evidence for an \spro\ dusting upon
a mostly \rpro\ enrichment pattern, but \citet{roederer10a} 
demonstrated that this interpretation is incorrect.
An \rpro\ enrichment alone is sufficient.
Consequently, the Sm and Eu isotopic fractions derived by \citet{roederer08a}
in \mbox{HD~175305}
should be interpreted as the isotopic fractions produced
by the \rpro\ \textit{in this case}, rather than the combined
yields of $s$- and \rpro\ nucleosynthesis.
Allowing for such variations could also inform the debate over the origin of 
the Ba isotopes in \mbox{HD~140283} 
\citep{magain95,lambert02,collet09,gallagher10}.

In the top panel of Figure~\ref{rprostars1}, we 
show a plot of the distribution of
the differences between the heavy element abundances and the S.S.\ 
\rpro\ residuals\footnote{
A number of nucleosynthetic processes contribute to the
production of the Sr-Y-Zr group, 
so the concept of \rpro\ residuals for Sr-Y-Zr is not appropriate.
The \rpro\ residuals are only used for an overall normalization
in Figure~\ref{rprostars1}.
See Section~\ref{hew} for a
fuller discussion of this point.}
for three stars (\mbox{CS~22892--052}, \mbox{HD~88609}, 
and \mbox{HD~122563}), based on Figure~5 of \citet{honda07}.
That study demonstrated clearly that the heavy element abundance pattern 
of the latter two stars
could not be matched by any combination of scaled S.S.\ 
\rpro\ or \spro\ patterns and was distinct from that of 
\mbox{CS~22892--052}.
In the bottom panel of 
Figure~\ref{rprostars1} we show a similar plot for 16 stars with
$-3.3 <$~[Fe/H]~$< -1.5$
(as well as \mbox{CS~22949--037} with [Fe/H]~$=-4.0$, \citealt{depagne02}).
In this panel, all abundance differences are normalized to Sr, the lightest
heavy element that is easily detectable in metal-poor stars.
To the best of our knowledge, 
these stars have not been enriched by the \spro.
The stars near the top of the diagram, with the smallest differences,
are those strongly enriched by the \rpro\ (e.g., 
\mbox{CS~22892--052}, \mbox{CS~31082--001}, and \mbox{HE~1523--0901}),
while the stars near the bottom of the diagram are those
deficient in the heavy elements
(e.g., \mbox{HD~88609} and \mbox{HD~122563}).
This reaffirms the conclusion of \citet{honda07} that 
the heavy element abundance pattern of \mbox{CS~22892--052}
is clearly distinct from that of either \mbox{HD~88609} or \mbox{HD~122563}.
The 13 other stars in Figure~\ref{rprostars1} 
appear to fill in the continuum between these two extremes.\footnote{
The lack of heavy elements ($Z >$~70) in the stars in the
bottom half of the bottom panel of
Figure~\ref{rprostars1} is due to both the overall weakness
of these species' lines and the lack of ultra-violet (UV) spectra
for all but one of these stars.
The resonance lines of several heavy \ncap\ species---including 
Lu~\textsc{ii} ($Z =$~71), Os~\textsc{ii} ($Z =$~76), 
Pt~\textsc{i} ($Z =$~78), Au~\textsc{i} ($Z =$~79), 
and Pb---are found in the near-UV.
}
While there is a considerable degree of scatter about the
mean difference from one element to the next in a single star,
the gross effect highlighted by Figure~\ref{rprostars1} is far beyond
any reasonable observational uncertainty.
This illustrates again one point made in Section~\ref{hew}:
\mbox{CS~22892--052} and \mbox{HD~122563} are not necessarily
archetypes of two distinct $r$-processes.
Rather, they may represent the extremes of a continuous range of
\rpro\ nucleosynthesis patterns---the full, main \rpro\ and 
an incomplete main \rpro---coupled with a CP nucleosynthesis component.

\subsection{Heavy Element Enrichment in Metal-Poor Globular Clusters}
\label{gc}

\begin{figure*}
\epsscale{0.85}
\begin{center}
\plotone{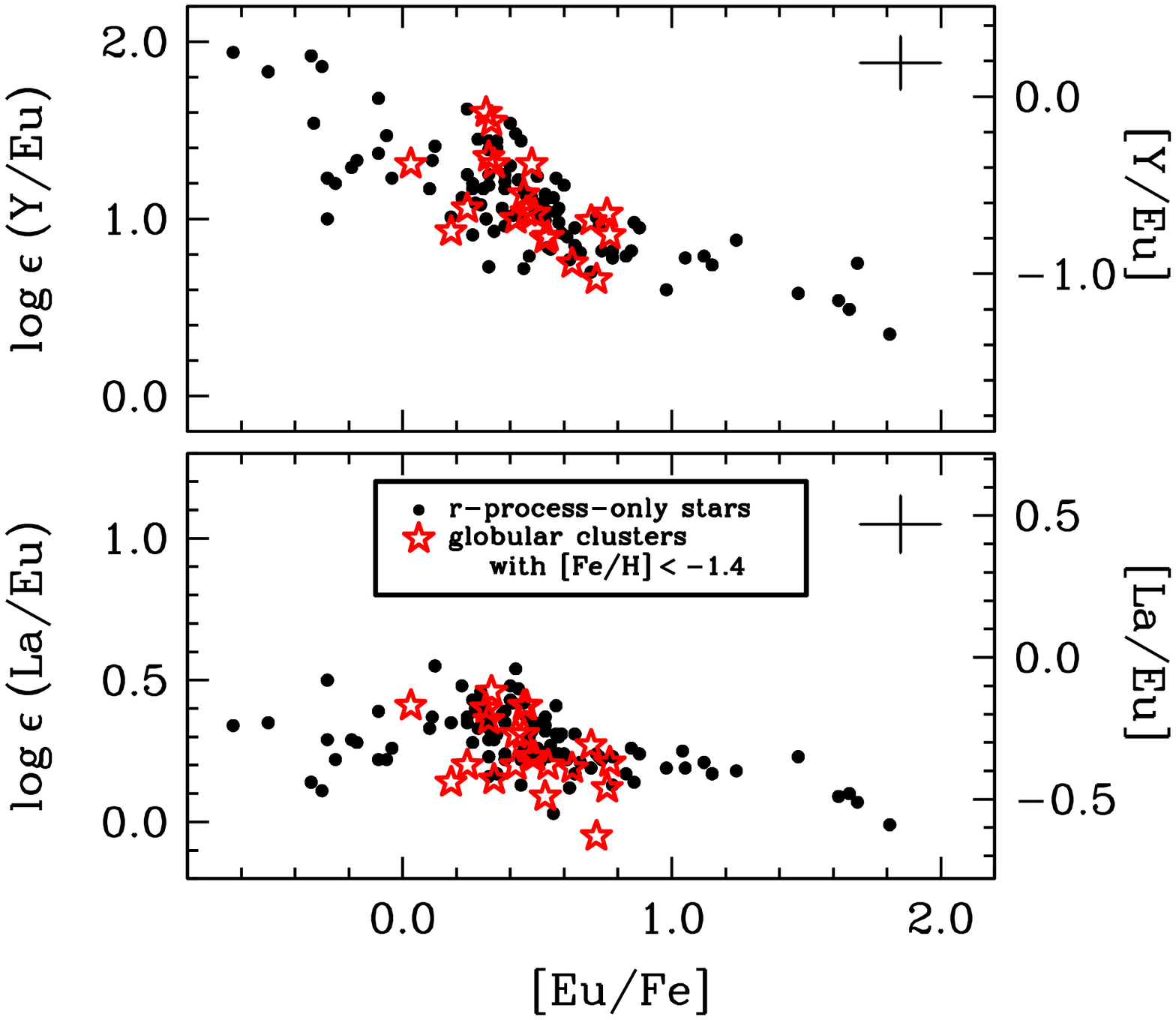}
\end{center}
\caption{
\label{gcplot}
Logarithmic Y/Eu and La/Eu ratios as a function of [Eu/Fe] for 
field stars and 25 Milky Way and LMC
globular clusters with [Fe/H]~$< -$1.4.
Only the cluster mean values are shown.
Here, ``\rpro-only'' denotes stars with 
[La/Eu]~$< 0.0$ (log~(La/Eu)~$< +$0.60; see Figure~\ref{simmererplot})
or otherwise classified as lacking
\spro\ material in Table~\ref{abundtab}.
A representative uncertainty is shown 
in the upper right corner of each panel.
}
\end{figure*}

In Figure~\ref{gcplot} we compare the mean logarithmic Y/Eu and La/Eu 
ratios between field stars and 25 metal-poor globular clusters 
([Fe/H]~$< -$1.4).
Most abundances are adopted from the compilation of \citet{pritzl05},
with original source references as follows:\footnote{
\citet{cavallo04} have derived La and Eu abundances for 8 giants in
globular cluster M80 ([Fe/H]~$= -$1.7), but \citet{lawler01} did
not report a log($gf$) value for the one line of La examined by 
\citet{cavallo04}, so we discard this cluster from our sample.}
Arp~2 \citep{mottini08},
M3 \citep{cohen05a},
M13 \citep{cohen05a},
M15 \citep{sobeck10},
M22 \citep{brown92,marino09},
M30 \citep{shetrone03},
M54 \citep{brown99},
M55 \citep{shetrone03},
M68 \citep{shetrone03},
M92 \citep{shetrone01},
NGC~2298 \citep{mcwilliam92},
NGC~3201 \citep{gonzalez98},
NGC~5694 \citep{lee06},
NGC~6287 \citep{lee02},
NGC~6293 \citep{lee02},
NGC~6397 \citep{norris95},
NGC~6541 \citep{lee02},
NGC~6752 \citep{yong05},
NGC~7492 \citep{cohen05b}, and
Pal~3 \citep{koch09}.
We also include 
5 clusters associated with the Large Magellanic Cloud (LMC;
2 from \citealt{johnson06} and 3 from \citealt{mucciarelli10}).

For \textit{all} clusters except one with [Fe/H]~$< -$1.4 and detected
Eu and either Y or La (or both), the relationships
between the cluster means fall exactly within the range set by
the field stars enriched by only \rpro\ material.
Intra-cluster star-to-star variations in globular cluster M15
also follow this relation \citep{otsuki06}.
Even in the one exception, NGC~2210 in the LMC, the [La/Eu]
ratio is $\sim$~0.2~dex \textit{lower} than the rest of the globular clusters
and field stars, indicating that the \spro\ could not have produced
these heavy elements.
Pb has only been detected in 4 stars in M13 and
5 stars in NGC~6752 by \citet{yong06}, 
but in these clusters it is clearly low,
$\langle$[Pb/Eu]$\rangle = -$0.73 and
$\langle$[Pb/Eu]$\rangle = -$0.48, respectively,
indicating that there has been no enrichment by the \spro.
Two metal-poor globular clusters in this sample 
are associated with the Sagittarius dwarf galaxy 
(Arp 1 and M54, \citealt{ibata95,law10}).
While the metal-rich stars in Sagittarius clearly have been 
enriched by the \spro\ (e.g., \citealt{chou10}), the metal-poor
stars and globular clusters appear to lack \spro\ material.
Pb, which has not been examined in any Sagittarius debris, 
would provide the strongest confirmation of this scenario.

While M54 likely formed elsewhere in Sagittarius and later migrated
to its center \citep{bellazzini08,carretta10}, 
M22 may itself be the nucleated core (i.e., the central
remnant after the outer layers have been stripped away)
of a dwarf spheroidal galaxy (dSph) 
like $\omega$~Centauri ($\omega$~Cen; \citealt{dacosta09}).
If so, 
it should not be unreasonable to expect chemical evolution in this system.
M22 shows an internal Fe spread, 
and the mean Y/Eu ratios are slightly different for the 
metal-rich and metal-poor stars in M22:
$\langle$[Y/Eu]$\rangle = -$0.4 for 
$\langle$[Fe/H]$\rangle = -$1.85~$\pm$~0.07 and
$\langle$[Y/Eu]$\rangle = $0.0 for 
$\langle$[Fe/H]$\rangle = -$1.62~$\pm$~0.06
%
(estimated from eight stars in Figure~21 of \citealt{marino09}).
Both [Y/Eu] ratios are well within the field star range in 
Figure~\ref{gcplot}.
Unlike the other clusters shown in Figure~\ref{gcplot}, however,
both groups of stars show an increase in [Y/Fe], [Ba/Fe], and [Nd/Fe]
without a corresponding increase in [Eu/Fe], indicating that 
\spro\ material is present in the metal-rich stars of M22.\footnote{
\citet{marino09} have no M22 stars in common with \citet{brown92},
and these studies used different Fe scales, so it is not obvious
whether the stars with low [La/Eu] derived by \citet{brown92}
belong to the metal-rich or metal-poor population.} 
While this cautions against a blanket \rpro\ interpretation for
the remaining clusters in Figure~\ref{gcplot}, 
to the best of our knowledge none of the other clusters (except M54)
show an internal metallicity spread and thus would not be expected
to show evolution in their heavy element ratios.
\citet{brown99} examined the heavy elements in only 5 stars at the 
peak of the metallicity distribution of M54, so the possibility of
heavy element evolution with metallicity is ripe for reexamination 
in this cluster.
In both M22 and M54, the Pb abundance would provide an unambiguous 
discriminant to test this hypothesis.

\subsection{The Appearance of $s$-process Material in the ISM}
\label{risespro}

Based on the observations displayed in Figure~\ref{simmererplot}, 
we suggested that the increase in La/Eu
with increasing [Fe/H] reflects the dispersion in \rpro\ nucleosynthesis 
rather than the onset of \spro\ enrichment in the ISM.
The Pb/Eu ratio, which should be a more robust indicator
of \spro\ enrichment, shows no upward trend in the metallicity
range $-$2.3~$<$~[Fe/H]~$< -$1.4 
(and no trend that
exceeds the minimum Pb/Eu ratio expected from AGB \spro\ production,
even if this ratio has been diluted by a factor of a few).
This result holds whether we consider only the stars marked by red circles
or all detections in stars with [Fe/H]~$< -$1.4.
The current observational data suggest 
that it is unlikely that the Pb in these stars
originated in the \spro, and \spro\ material
does not seem to have been dispersed throughout 
the ISM until the mean metallicity exceeds
at least [Fe/H]~$= -$1.4.
This is in agreement with previous investigations 
that used other tracers of AGB enrichment
(e.g., \citealt{melendez07}).

Other studies have demonstrated a clear onset of the \spro\ in 
globular clusters with multiple stellar populations that
may be nucleated cores of dSphs.
These clusters include $\omega$~Cen and M22.
In $\omega$~Cen, many stars with [Fe/H]~$\gtrsim -$1.6 show 
[La/Eu]~$>$~0.0, our minimum AGB discriminant, indicating that
this increase in [La/Eu] is not due to a dispersion
in the \rpro\ ratios 
(\citealt{johnson10} and references therein).
\citet{marino09} and \citet{dacosta09} have demonstrated
that M22 resembles $\omega$~Cen in that it
shows an analogous increase in [Ba/Fe] and [Nd/Fe] 
as metallicity increases from [Fe/H]~$\sim -$1.8.
According to the Pb/Eu ratios in our stellar sample,
the onset of the \spro\ occurs at
a higher mean metallicity in the halo field stars of the Milky Way
than in $\omega$~Cen or M22.
If star formation proceeded at a higher rate in the Milky Way
than in dwarf galaxies or their former nuclei,
it would be very surprising
if \spro\ material produced by AGB stars should have been 
dispersed throughout the ISM
of the Milky Way at a metallicity significantly lower than in the 
dSphs.

\subsection{The Ubiquity of $r$-process Material in Metal-Poor Stars}
\label{enrichment}

In many metal-poor stars the absorption lines of the heavy elements
are so weak that only Sr and Ba may be detected. 
Figure~7 of \citet{sneden08} shows the range of [Ba/Sr] ratios
observed in metal-poor stars as a function of [Ba/Fe];
these ratios are analogous to the 
[Y/Eu] and [Eu/Fe] ratios shown in our Figure~\ref{resultplot1}.
The low-C stars in their plot ([C/Fe]~$< +$0.25) likely do 
not contain significant amounts of \spro\ material, and
yet they span ranges of $-$1.5~$<$~[Ba/Sr]~$< +$0.6
and $-$2.0~$<$~[Ba/Fe]~$< +$1.0.  
The trends between Fe, Sr, and Ba are similar to those in Fe, Y, and Eu.
At extremely low metallicities
the Ba and Sr seem to indicate that in most cases
the relationship between the light and heavy \ncap\ elements holds.
It is logical to assume (based on the arguments in Section~\ref{risespro}
and the available observational data) that most of
these stars are enriched by the \rpro.
Thus it is plausible that the nucleosynthesis mechanisms described
in Section~\ref{hew} may operate at metallicities at least as low as
[Fe/H]~$\sim -$4.0 (see \citealt{mcwilliam98}, \citealt{honda04},
\citealt{francois07}, and \citealt{lai08}).

How frequently are heavy elements found in metal-poor stars?
Can Sr and Ba always be detected if Mg can be detected?
\citet{barklem05} performed an abundance analysis on
a sample of 253 metal-poor field stars with a range of metallicities 
($-$3.8~$<$~[Fe/H]~$< -$1.5), effective temperatures 
(4300~$< T_{\rm eff} <$~6800~K, mostly giants and subgiants), 
and distances (most with 1~$\lesssim D \lesssim$~10~kpc).
Stars with strong molecular C features or double-lined spectroscopic
binaries were deliberately excluded from their sample, so the
majority of the 253 stars should not be significantly enriched
in \spro\ material.
\citet{barklem05} derived Mg abundances for 245 stars (97\%),
Sr abundances for 245 stars (97\%), and 
Ba abundances for 220 stars in their sample (87\%).
The fraction of stars with detected Mg, Sr, and Ba increases to 
100\%, 99\%, and 92\% if only those stars with $T_{\rm eff} <$~5500~K
are considered (159~stars total), and 
all increase to 100\% if only stars with $T_{\rm eff} <$~4800~K
are considered (34~stars).
Thus it would seem that the occasional non-detection of Sr and Ba
can be attributed to the strength of these elements' lines relative
to the continuous opacity that increases with increasing $T_{\rm eff}$.

\begin{figure}
  \epsscale{1.15}
\plotone{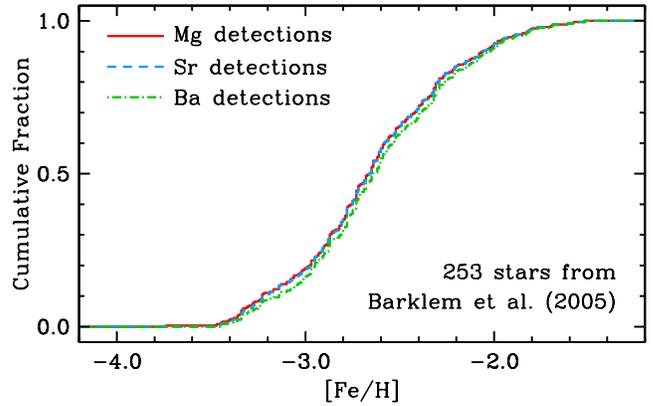}
\caption{
\label{barkplot}
Cumulative distributions of stars with Mg, Sr, or Ba detections
as a function of [Fe/H] for 253 stars from the sample of \citet{barklem05}.
}
\end{figure}

To investigate further,
in Figure~\ref{barkplot} we plot the cumulative distributions
of these stars (at all values of $T_{\rm eff}$) as a function of [Fe/H].
These distributions are very similar.
A Kolmogorov-Smirnov test confirms that each of the distributions 
of Mg and Sr or Mg and Ba are not significantly different
at the 99.5\% confidence level.
This result is unchanged if the distributions are considered
a function of $T_{\rm eff}$ and is insensitive to 
whether the $s$ and $r+s$ stars
listed in Table~8 of \citet{jonsell06} are included.
Thus we conclude that Sr and Ba are 
present in nearly all metal-poor field stars.\footnote{
A few stars, such as Draco~119 ([Fe/H]~$= -$3.0; \citealt{fulbright04}), 
do not appear to have any significant accumulation 
of elements heavier than the Fe-group
([Sr/Fe]~$< -$2.5, [Ba/Fe]~$< -$2.6).
This particular star also has $\alpha$/Fe ratios different
from most other metal-poor stars in the halo
([Mg/Fe]~$= +$0.5, [Si/Fe]~$< +$0.2, [Ca/Fe]~$= -$0.1), 
and it is likely that the SN that enriched Draco~119 is 
different than those that enriched the majority of metal-poor field stars.}
If detectable quantities of \spro\ material 
are not widespread in the ISM 
at [Fe/H]~$< -$1.4, then it seems that these stars have been
enriched by the \rpro\ and associated
CP nucleosynthesis (cf.\ \citealt{truran81}).

\subsection{Implications for Chemical Evolution: \\
Mixing or Variable Heavy Element Yields?}

The chemical composition of our sample of \rpro-only stars
can be summarized as follows.
Numerous previous studies have shown that 
the $\alpha$ elements correlate strongly with Fe in these stars, 
typically [$\alpha$/Fe]~$\sim +$0.2 to 0.5. 
Figure~\ref{resultplot1} demonstrates that there is also a 
correlation between [Y/Eu] and [Eu/Fe] in these stars
(analogous to the correlation between [Sr/Ba] and [Ba/Fe]
shown in Figure~7 of \citealt{sneden08}).
Most metal-poor stars with [Fe/H]~$\lesssim -$1.4
whose atmospheres retain a fossil record of their birth composition
follow this relationship, with an intrinsic
scatter of a factor of $\sim$~2--8 in Y/Eu that is much
smaller than the factor of $>$~30 over which the relationship extends.
Nearly all of these stars
contain detectable amounts of elements with $A >$~130.

The $\alpha$ and Fe-group elements were likely 
produced in Type~II SNe.
It is unlikely that a relationship would exist among the
heavy elements Y (or Sr) and Eu (or Ba) if these elements originated in 
separate or uncorrelated events (see also \citealt{johnson02b}).
This suggests that mixing between heavy element patterns 
(typified as the extreme cases 
\mbox{CS~22892--052} and \mbox{HD~122563}) 
is not alone responsible
for the range of Y/Eu or Sr/Ba ratios observed in the atmospheres
of metal-poor stars.
Furthermore, the relationship between [Y/Eu] and [Eu/Fe] suggests that most
metal-poor stars have not been severely diluted with Fe from other events
that did not produce elements heavier than the Fe-group.\footnote{
One star, UMi~COS82 ($=$~UMi~199) 
with [Fe/H]~$= -$1.42 and [Eu/Fe]~$= +$1.24,
appears to be an exception.
A number of observational studies have shown that 
the metal-rich stars of the Ursa Minor dSph
galaxy, including COS82,
have been enriched by the products of both Type~II and Type~Ia
SNe, as demonstrated by their reduced (relative to stars in the
Milky Way halo at the same metallicities) [Si/Fe], [Ca/Fe], and [Ti/Fe]
ratios \citep{shetrone01,sadakane04,cohen10}.
This implies that the [Eu/Fe] ratio in the Type~II contribution
to the gas from which COS82 would eventually form may have
been higher before additional Fe was added from the Type~Ia contribution.
This could explain the apparent enhancement of COS82 in the [Y/Fe] 
and [Eu/Fe] ratios in Figure~\ref{znyplot} and Figure~\ref{eufeplot}, 
respectively, relative to other stars at [Fe/H]~$= -$1.4.}
If we accept that the Y (or Sr) and Eu (or Ba) 
in these stars are produced by the \rpro\
and associated CP nucleosynthesis,
the simplest explanation for the ubiquitous presence 
of Sr and Ba described in Section~\ref{enrichment} 
is that the \rpro\ might also be associated with Type~II SNe. 

As a reminder, our use of the term ``\rpro\ enriched star''
in this context refers to those stars that contain at least 
a detectable trace of Ba and possibly heavier elements.
If we assume that in the absence of \spro\ enrichment only
the \rpro\ is capable of producing detectable quantities 
of these heavy elements (i.e., they were not produced in a
CP process), then the currently-available
observational data suggest that the \rpro\ could be a common feature of
nearly all Type~II SN events.
Of course not all \rpro\ events will produce yields that
enrich the next generation of stars to resemble
\mbox{CS~22892--052} or \mbox{CS~31082--001} 
(with [Eu/Fe]~$= +$1.6), but
even stars with solar [Eu/Fe] ratios can have a near-perfect
match between their REE abundances and the REE abundances
in the strongly-enriched \rpro\ stars 
(see, e.g., Figure~12 of \citealt{roederer10a}).
It is reasonable to suppose that the majority of \rpro\ events will
enrich the next generation of stars with small amounts of
\rpro\ material. 
We caution that
the \rpro-only sample shown in Figure~\ref{resultplot1} is strongly
biased towards $r$-rich stars and should not be taken as an
estimate of the \rpro\ yield distribution function.

Conditions within the SN wind may be variable, and therefore
the \rpro\ yields will be also (perhaps due to progenitors
of different mass ranges, e.g., \citealt{arnone05}), 
but it seems reasonable to conclude
that some heavy elements are produced in nearly 
all Type~II SN events.
The variable ratio of \rpro\ to CP yields---the
``strength'' of the \rpro---provides a natural explanation for
the large dispersion in [Eu/Fe] ratios around [Fe/H]~$\sim -$3.0
(see Figure~\ref{eufeplot}).
The decreased dispersion in [Eu/Fe]
with increasing [Fe/H] may reflect the growing chemical homogeneity
of the ISM.
If so, significant mixing in the halo (progenitors'?)
ISM had already occurred long before the mean metallicity at which
Type~Ia SNe or low-metallicity intermediate-mass AGB stars 
began contributing significant amounts of material to the ISM.

Type~II SNe alone may be capable of 
producing the diversity of heavy element abundances observed in stars
at [Fe/H]~$\lesssim -$3.0.  
Thus there is no reason to exclude the possibility that 
some stars at these metallicities may have been
enriched by a very small number of SNe, perhaps even one.
For example, \citet{simon10} have analyzed the abundance pattern 
of a star with [Fe/H]~$= -$3.2 in the low luminosity 
dwarf galaxy Leo~IV;
the abundance pattern in this star is consistent with other 
metal-poor field stars (i.e., $\alpha$-enhanced, etc.),
and it does contain very low but detectable traces of Sr and Ba.
This star is not strongly enriched in C ([C/Fe]~$< -$0.1), 
suggesting that the Ba was produced by an incomplete main \rpro.
Given the overall low luminosity and metallicity of Leo~IV,
a very small \textit{total} Fe abundance is present in the entire galaxy
(consistent with that produced by a single SN event), and this led
\citet{simon10} to hypothesize that a single SN may have
enriched Leo~IV if metals were not lost from the galaxy by winds.
We encourage efforts to demonstrate that more metal-poor stars were---or were 
not---enriched by the yields of a single SN event, for this
could place very strong constraints on the nature of the
explosion and nucleosynthesis mechanisms.
Furthermore, since the CP and \rpro\ yields seem to vary much more than
the $\alpha$ or Fe-group yields, these heavy elements may be a more sensitive
probe of the nature of the SN progenitor than the lighter elements are.

\section{Conclusions}
\label{conclusions}

We have compiled a sample of 161 metal-poor stars with 
$-$4.2~$<$~[Fe/H]~$< -$0.6.
These stars include detections or upper limits for
Zn, Y, La, Eu, or Pb, including 
abundances or upper limits for Pb in 120 stars.
New Zn, Y, and Pb
abundances are derived from the high-resolution, high-S/N
spectra described in \citet{simmerer04}
or are compiled from the literature.
From this sample we identify a subset of stars
that has not been enriched by the \spro,
and we characterize the heavy element enrichment patterns in this subset.
Based on the observational data available at present, 
our main conclusions can be summarized as follows.

(1) 
The Pb/Eu ratio 
can be used to successfully identify metal-poor
stars that lack any detectable trace of the \spro.
At low metallicity the \spro\ produces large amounts of 
Pb relative to, e.g., Fe and Eu, and high Pb/Fe or Pb/Eu ratios
are clear observational signatures of the \spro\ in metal-poor stars.
Based on models of \spro\ nucleosynthesis in intermediate mass
stars on the AGB, the minimum \spro\ ratios predicted 
([Pb/Eu]~$= +$0.3)
can be used to identify stars that have not been enriched
by the \spro.

(2)
The relationship between the light (e.g., Sr, Y, and Zr) and heavy
(e.g., Ba, La, Eu, and heavier) \ncap\ material produced by the 
\rpro\ can be characterized based on observations of metal-poor stars.
Stars strongly enriched by the \rpro, such as \mbox{CS~22892--052},
are overabundant in the heavy elements relative to the light ones, and 
stars such as \mbox{HD~122563} are deficient in the heavy elements
(rather than overabundant in the light ones).
We have culled our sample of stars that show evidence of \spro\ 
enrichment, and the data for the remaining $r$-only stars suggest
that these two stars are not archetypes of two distinct $r$-processes, but
rather they may represent the extremes of a continuous range of
\rpro\ nucleosynthesis patterns.

(3)
We identify a dispersion of abundance ratios among the 
rare earth elements produced in the \rpro.
This dispersion spans a range of at least 
$+$0.0~$\lesssim$~log~(La/Eu)~$\lesssim +$0.5
(or $-$0.6~$\lesssim$~[La/Eu]~$\lesssim -$0.1)
which cautions
against using the La/Eu ratio (or similar ratios, e.g., Ba/Eu, or
isotopic fractions, e.g., the Ba, Sm, or Eu isotopes)
alone as a precision discriminant of 
$s$- and \rpro\ nucleosynthesis contributions to a given star.

(4)
The ranges in Y/Eu and La/Eu can be reproduced by 
nucleosynthesis predictions from simulations of the high-entropy neutrino
wind (HEW) of a core-collapse SN.
In these simulations the strength of the $r$-process 
(the ratio Y$_n$/Y$_{\rm seed}$)
is determined by the entropy, the electron abundance,
and the expansion velocity of matter in the SN.
The $\alpha$-rich freeze-out and $\beta$-delayed neutron recapture processes 
produce an abundance pattern for the Sr-Y-Zr group
that fits the requirements for the light element primary process (LEPP).
Conditions consistent with the traditional understanding of
an \rpro\ (e.g., 10$^{23} \leq$~n$_{n} \leq 10^{28}$)
are required to produce detectable
amounts of material heavier than the 2nd \rpro\ peak, but 
these conditions themselves do not produce 
significant quantities of the lighter elements.
This result reaffirms earlier suggestions
that multiple
processes (besides the \spro) or a diversity of physical conditions
in the \rpro\ must contribute to the nucleosynthesis of the 
Sr-Y-Zr group; thus, simple \rpro\ residuals 
($N_{\rm S.S., r} \equiv N_{\rm S.S., total} - N_{\rm S.S., s}$)
are inadequate descriptions of the origins of these nuclei.

(5)
For the full sample of stars with [Fe/H]~$< -1.4$,
the [Pb/Eu] ratios show no significant increase 
with increasing [Fe/H], and a number of stars with 
[Pb/Eu]~$\leq -$0.7 have metallicities as high as
[Fe/H]~$= -$1.4.
These observations might suggest
that \spro\ material produced in intermediate-mass stars on the AGB
is not widespread in the ISM until the overall
Galactic metallicity grew considerably, perhaps even to 
[Fe/H]~$= -$1.4.
We cannot make any definitive statements about the \spro\
at higher metallicities from this sample.
The heavy elements in 
most stars with [Fe/H]~$< -$1.4 that have not received \spro\ material 
directly from an AGB binary companion
appear to have been produced by the \rpro\ 
(and the associated CP process).

(6)
This \rpro\ enrichment pattern is common to both field stars and
metal-poor globular clusters.
Except for M22 and M54, the heavy \ncap\ elements in the 
23 other metal-poor
([Fe/H]~$< -$1.4) globular clusters examined here 
seem to have been produced by the \rpro\ 
(and the associated CP process), 
and the globular clusters follow the same
\rpro\ trends observed in metal-poor field stars.
Based on the currently-available observational evidence, 
these 23 metal-poor ([Fe/H]~$< -$1.4) Milky Way and LMC globular clusters
have \ncap\ abundance ratios that suggest significant amounts of \spro\ 
material were not present in the ISM from which these cluster stars formed.

(7)
At least small amounts of material heavier than the Fe-group have been 
detected in nearly all metal-poor stars.
The light element abundance patterns 
(i.e., among the $\alpha$ and Fe-group elements, 8~$\leq Z \leq$~32)
in these stars are associated with Type~II core-collapse SNe.
The simplest explanation for
the ubiquitous presence of Sr and Ba in these stars is that
the nucleosynthesis mechanisms described by our HEW model
($\alpha$-rich freeze-out, $\beta$-delayed neutron emission
and recapture, and the \rpro)
are also present in core-collapse SNe,
and at least one of these mechanisms is in operation 
in nearly all core-collapse SN events.
In this scenario, the scatter in \ncap-to-Fe ratios at [Fe/H]~$\lesssim -$3.0
may be attributed to differing strengths of \rpro\ events
rather than infrequent occurrences of \rpro\ events.
The \rpro\ is not a rare phenomenon: 
nearly all normal metal-poor stars have been enriched by the \rpro.

\acknowledgments

We thank 
A.\ Frebel and J.\ Sobeck
for providing abundance derivations in advance of publication, 
E.\ Robinson for helpful discussions, 
A.\ Frebel, G.\ Preston, and M.\ Shetrone 
for insightful comments on early versions of the manuscript,
and the referee for providing a number of suggestions that
have improved the presentation of this report.
This research has made use of the 
NASA Astrophysics Data System (ADS),
NIST Atomic Spectra Database, and the
SIMBAD database (operated at CDS, Strasbourg, France),
and the Milky Way Spheroid Substructure database.\footnote{
http://www.rpi.edu/\~{}newbeh/mwstructure/MilkyWaySpheroidSubstructure.html}
Funding for this project has been generously provided by 
the U.~S.\ National Science Foundation
(grants AST~07-07447 to J.C. and AST~09-08978 to C.S.).
M.L.\ is supported by a Monash Research Fellowship.

{\it Facilities:} 
\facility{Smith (2dCoude)}

\begin{center}

\begin{deluxetable}{lcccccclccccc}
\tablecaption{Equivalent Widths
\label{ewtab}}
\tablewidth{0pt}
\tabletypesize{\scriptsize}
\tablehead{
\colhead{Star} &
\colhead{Zn~\textsc{i}} &
\colhead{Zn~\textsc{i}} &
\colhead{Y~\textsc{ii}} &
\colhead{Y~\textsc{ii}} &
\colhead{Y~\textsc{ii}} &
\colhead{} &
\colhead{Star} &
\colhead{Zn~\textsc{i}} &
\colhead{Zn~\textsc{i}} &
\colhead{Y~\textsc{ii}} &
\colhead{Y~\textsc{ii}} &
\colhead{Y~\textsc{ii}} \\
\colhead{} &
\colhead{4722\AA} &
\colhead{4810\AA} &
\colhead{4833\AA} &
\colhead{5087\AA} &
\colhead{5200\AA} &
\colhead{} &
\colhead{} &
\colhead{4722\AA} &
\colhead{4810\AA} &
\colhead{4833\AA} &
\colhead{5087\AA} &
\colhead{5200\AA} \\
\colhead{} &
\colhead{(m\AA)} &
\colhead{(m\AA)} &
\colhead{(m\AA)} &
\colhead{(m\AA)} &
\colhead{(m\AA)} &
\colhead{} &
\colhead{} &
\colhead{(m\AA)} &
\colhead{(m\AA)} &
\colhead{(m\AA)} &
\colhead{(m\AA)} &
\colhead{(m\AA)} }
\startdata
BD$-$01~0306    &   38.7 &  45.9 &  34.4 &  24.0 &  14.7  & &   HD~25532        &   49.4 &  57.9 &  74.6 &  54.7 &  40.8 \\
BD$-$01~2582    &   15.5 &  17.6 &  31.0 &  20.4 &  11.6  & &   HD~26297        &   46.6 &  52.7 &  70.8 &  55.9 &  47.2 \\
BD$+$19~1185    &   28.7 &  33.7 &  25.1 &  13.7 &  11.8  & &   HD~29574        &   49.9 &  51.2 &  97.0 &  76.1 &  68.2 \\
BD$+$52~1601    &   65.4 &  69.4 &  78.5 &  62.1 &  50.9  & &   HD~37828        &   57.6 &  62.6 &  94.3 &  74.4 &  71.7 \\
G005-001        &   31.0 &  38.0 &  20.9 &  11.7 &   7.2  & &   HD~44007        &   39.6 &  46.0 &  58.0 &  43.0 &  34.3 \\
G009-036        &   27.7 &  28.6 &  25.0 &  25.0 &   9.2  & &   HD~63791        &   43.0 &  50.6 &  61.1 &  46.8 &  36.3 \\
G017-025        &   29.3 &  34.7 &  34.6 &  19.4 &\nodata & &   HD~74462        &   54.2 &  58.2 &  70.8 &  55.5 &  46.0 \\
G023-014        &   32.5 &  38.8 &  29.7 &  26.4 &  19.1  & &   HD~82590        &   19.5 &  25.3 &  48.0 &  28.3 &  17.4 \\
G028-043        &   13.7 &  19.7 &  16.5 &   5.8 &\nodata & &   HD~85773        &   46.1 &  52.4 &  40.5 &  25.3 &\nodata\\
G029-025        &   45.0 &  47.6 &  35.9 &  22.3 &  18.1  & &   HD~88609        &    9.3 &  14.9 &  23.0 &\nodata&   8.2 \\
G040-008        &   44.7 &  50.5 &  31.6 &  19.2 &  16.6  & &   HD~101063       &   41.9 &  47.7 &  51.7 &  36.5 &  29.6 \\
G058-025        &   17.4 &\nodata&  19.4 &  10.5 &   6.2  & &   HD~103036       &   68.0 &  74.0 & 123.4 &  97.1 &  87.0 \\
G059-001        &   43.8 &  48.9 &  34.5 &  18.9 &  17.6  & &   HD~103545       &   17.4 &  23.2 &  32.6 &  22.3 &  12.4 \\
G063-046        &   48.0 &  54.4 &  38.0 &  27.9 &  18.7  & &   HD~105546       &   46.6 &  57.4 &  61.3 &  47.8 &  37.1 \\
G068-003        &   61.6 &  64.4 &  53.8 &  40.1 &  36.4  & &   HD~105755       &   54.8 &  59.7 &  42.0 &  29.4 &  19.8 \\
G074-005        &   36.0 &  43.5 &  25.4 &  15.5 &  11.1  & &   HD~106516       &   42.2 &  47.9 &  32.5 &  23.7 &  13.1 \\
G090-025        &   11.1 &  14.9 &   9.1 &   3.7 &\nodata & &   HD~107752       &   13.0 &  11.6 &  16.2 &  10.7 &\nodata\\
G095-057A       &   36.9 &  43.4 &  47.9 &  31.1 &\nodata & &   HD~108317       &   12.8 &  16.2 &  17.8 &   8.9 &   4.2 \\
G095-057B       &   33.7 &  38.4 &\nodata&  29.0 &\nodata & &   HD~110184       &   29.4 &  33.4 &  59.5 &  42.4 &  32.7 \\
G102-020        &   29.7 &  34.6 &  22.3 &  12.6 &\nodata & &   HD~115444       &    6.7 &  10.2 &  13.9 &   8.5 &   4.3 \\
G102-027        &   66.7 &  70.2 &  56.1 &  42.3 &  35.0  & &   HD~121135       &   60.9 &  66.9 &  78.5 &  57.7 &  44.5 \\
G113-022        &   40.1 &  44.8 &  47.0 &  33.9 &  25.3  & &   HD~122563       &   13.5 &  19.6 &  24.2 &  14.0 &   6.9 \\
G122-051        &   19.7 &  25.1 &  21.6 &   9.4 &\nodata & &   HD~122956       &   46.0 &  51.3 &  66.5 &  48.2 &  42.3 \\
G123-009        &   28.5 &  32.6 &  31.4 &  18.3 &  14.0  & &   HD~124358       &   37.0 &  43.5 &  48.9 &  30.9 &  21.2 \\
G126-036        &   37.9 &  44.8 &  45.2 &  34.5 &  26.3  & &   HD~132475       &   23.0 &  29.7 &  32.7 &  20.8 &  13.9 \\
G126-062        &   10.5 &  16.1 &  12.4 &   5.4 &   3.7  & &   HD~135148       &   55.7 &  55.5 &  96.1 &  67.9 &  59.8 \\
G140-046        &   35.6 &  41.7 &  54.2 &  33.3 &\nodata & &   HD~141531       &   48.7 &  55.5 &  80.0 &  61.4 &  53.1 \\
G153-021        &   57.6 &  62.1 &  43.3 &  38.4 &  19.0  & &   HD~166161       &   61.4 &  67.9 &  87.3 &  69.6 &  56.0 \\
G176-053        &   19.5 &  22.9 &  14.4 &   8.6 &   5.3  & &   HD~171496       &   78.9 &  80.5 &  81.0 &  68.0 &  60.4 \\
G179-022        &   38.4 &  47.5 &  47.0 &  32.9 &  25.3  & &   HD~184266       &   26.2 &  35.0 &  52.6 &  29.6 &  13.8 \\
G180-024        &   15.9 &  23.2 &  16.1 &   9.8 &   5.8  & &   HD~186478       &   19.6 &  24.8 &  40.4 &  27.5 &  16.8 \\
G188-022        &   19.7 &  28.2 &  28.2 &  16.2 &  10.3  & &   HD~187111       &   50.5 &  55.2 &  80.4 &  64.9 &  59.1 \\
G191-055        &    8.0 &  11.9 &   6.2 &\nodata&\nodata & &   HD~188510       &   13.2 &  18.4 &   9.7 &   6.3 &\nodata\\
G192-043        &   14.6 &  17.2 &  13.8 &   7.9 &\nodata & &   HD~193901       &   25.0 &  31.3 &  21.0 &  11.6 &   7.1 \\
G221-007        &   44.3 &  49.3 &  42.8 &  32.5 &  29.2  & &   HD~194598       &   23.7 &  28.2 &  21.4 &  14.2 &   7.2 \\
HD~2665         &   19.3 &  25.4 &  19.8 &  10.8 &   5.8  & &   HD~201891       &   30.5 &  36.4 &  23.5 &  13.5 &   7.7 \\
HD~3008         &   51.4 &  53.3 &  83.0 &  66.0 &  54.5  & &   HD~206739       &   49.1 &  55.2 &  69.8 &  52.6 &  45.7 \\
HD~6755         &   28.7 &  36.7 &  36.1 &  25.5 &  17.0  & &   HD~210295       &   56.0 &  60.3 &  73.2 &  56.8 &  48.8 \\
HD~6833         &   63.6 &  64.8 &  85.9 &  67.8 &\nodata & &   HD~214362       &   14.2 &  18.6 &  37.0 &  20.6 &   8.2 \\
HD~21581        &   38.9 &  45.2 &  54.3 &  41.5 &  32.3  & &   HD~218857       &   21.2 &  30.6 &  22.5 &  14.2 &   8.5 \\
HD~23798        &   35.2 &  39.9 &  70.4 &  52.3 &  45.6  & &   HD~233666       &   33.9 &  42.1 &  48.4 &  36.2 &  21.9 \\
HD~25329        &   12.4 &  14.0 &  23.5 &  10.9 &\nodata & &                   &        &       &       &       &       \\
\enddata
\end{deluxetable}

\clearpage
\LongTables

\begin{deluxetable}{lccccccccc}
\tablecaption{Stellar Abundances
\label{abundtab}}
\tablewidth{0pt}
\tabletypesize{\scriptsize}
\tablehead{
\colhead{Star} &
\colhead{[Fe/H]} &
\colhead{Ref.} &
\colhead{\eps{Zn~\textsc{i}}} &
\colhead{\eps{Y~\textsc{ii}}} &
\colhead{\eps{La~\textsc{ii}}} &
\colhead{\eps{Eu~\textsc{ii}}} &
\colhead{\eps{Pb~\textsc{i}}} &
\colhead{Ref.} &
\colhead{Class\tablenotemark{a}} }
\startdata
BD$-$01~0306    &   $-$1.13 & 29 &  $+$3.57 &   $+$1.05 &  $+$0.12 &  $-$0.27 &  $< +$1.20   & 29, 36 & 0 \\
BD$-$01~2582    &   $-$2.21 & 29 &  $+$2.51 &   $+$0.17 &  $-$0.05 &  $-$1.03 &    $+$0.77   & 29, 36 & 0 \\
BD$-$18~5550    &   $-$3.05 & 20 &  $+$1.94 &   $-$1.81 &  $-$2.52 &  $-$2.81 &    \nodata   & 20     & 0 \\
BD$+$01~2916    &   $-$1.92 & 2  &  \nodata &   $-$0.21 &  $-$0.87 &  $-$1.22 &    $-$0.20   & 2, 4   & 1 \\
BD$+$04~2621    &   $-$2.52 & 20 &  \nodata &   $-$0.69 &  $-$2.29 &  $-$2.63 &    \nodata   & 20     & 0 \\
BD$+$06~0648    &   $-$2.14 & 2  &  \nodata &   $-$0.09 &  $-$0.95 &  $-$1.50 &  $< +$0.00   & 2, 4   & 1 \\
BD$+$08~2856    &   $-$2.12 & 20 &  $+$2.59 &   $-$0.15 &  $-$1.03 &  $-$1.16 &    \nodata   & 20     & 0 \\
BD$+$10~2495    &   $-$2.31 & 26 &  $+$2.32.&   $-$0.52 &  $-$1.36 &  $-$1.69 &  $< +$0.88   & 26     & 2 \\
BD$+$17~3248    &   $-$2.08 & 8  &  $+$2.58 &   $+$0.04 &  $-$0.55 &  $-$0.78 &  $< +$0.27   & 8, 25  & 1 \\
BD$+$19~1185    &   $-$1.09 & 29 &  $+$3.38 &   $+$0.85 &  $+$0.06 &  $-$0.23 &    $+$0.77   & 29, 36 & 0 \\
BD$+$29~2356    &   $-$1.59 & 26 &  $+$3.05 &   $+$0.52 &  $-$0.47 &  $-$0.69 &    $+$0.35   & 26     & 2 \\
BD$+$30~2611    &   $-$1.50 & 26 &  $+$2.87 &   $+$0.41 &  $-$0.24 &  $-$0.36 &    $+$0.56   & 26     & 2 \\
BD$+$52~1601    &   $-$1.40 & 29 &  $+$3.35 &   $+$0.71 &  $-$0.13 &  $-$0.51 &  $< +$0.60   & 29, 36 & 0 \\
BS~16477--003   &   $-$3.36 & 6  &  $+$1.42 &   $-$1.20 &  \nodata &  \nodata &    \nodata   & 6, 10  & 0 \\
BS~17569--049   &   $-$2.88 & 6  &  $+$1.95 &   $-$0.63 &  $-$1.40 &  $-$1.64 &    \nodata   & 6, 10  & 0 \\
CD$-$36~1052    &   $-$1.79 & 26 &  $+$2.88 &   $+$0.34 &  $-$0.40 &  $-$0.82 &  $< +$2.47   & 26     & 2 \\
CD$-$38~0245    &   $-$4.19 & 6  &  $+$1.10 &   $-$2.43 &  \nodata &  \nodata &    \nodata   & 6, 10  & 0 \\
CS~22169--035   &   $-$3.04 & 6  &  $+$1.66 &   $-$1.21 &  \nodata &  \nodata &    \nodata   & 6, 10  & 0 \\
CS~22172--002   &   $-$3.86 & 6  &  $+$1.23 &   $-$2.63 &  \nodata &  \nodata &    \nodata   & 6, 10  & 0 \\
CS~22186--025   &   $-$3.00 & 6  &  $+$1.92 &   $-$1.10 &  $-$1.71 &  $-$1.94 &    \nodata   & 6, 10  & 0 \\
CS~22189--009   &   $-$3.49 & 6  &  $+$1.57 &   $-$2.11 &  \nodata &  \nodata &    \nodata   & 6, 10  & 0 \\
CS~22873--055   &   $-$2.99 & 6  &  $+$1.87 &   $-$1.31 &  $-$2.36 &  $-$2.64 &    \nodata   & 6, 10  & 0 \\
CS~22873--166   &   $-$2.97 & 6  &  $+$1.81 &   $-$0.89 &  $-$2.64 &  $-$2.75 &    \nodata   & 6, 10  & 0 \\
CS~22878--101   &   $-$3.25 & 6  &  $+$1.75 &   $-$1.32 &  $-$2.57 &  $-$2.79 &    \nodata   & 6, 10  & 0 \\
CS~22891--209   &   $-$3.29 & 6  &  $+$1.76 &   $-$1.18 &  $-$2.47 &  $-$2.86 &    \nodata   & 6, 10  & 0 \\
CS~22892--052   &   $-$3.10 & 30 &  $+$1.59 &   $-$0.42 &  $-$0.87 &  $-$0.96 &  $< -$0.15   & 25, 30 & 3 \\
CS~22896--154   &   $-$2.69 & 6  &  $+$2.17 &   $-$0.33 &  $-$1.17 &  $-$1.31 &    \nodata   & 6, 10  & 0 \\
CS~22897--008   &   $-$3.41 & 6  &  $+$1.86 &   $-$1.08 &  \nodata &  \nodata &    \nodata   & 6, 10  & 0 \\
CS~22948--066   &   $-$3.14 & 6  &  $+$1.83 &   $-$1.98 &  \nodata &  \nodata &    \nodata   & 6, 10  & 0 \\
CS~22952--015   &   $-$3.43 & 6  &  $+$1.42 &   $-$2.12 &  \nodata &  \nodata &    \nodata   & 6, 10  & 0 \\
CS~22953--003   &   $-$2.84 & 6  &  $+$1.91 &   $-$0.49 &  $-$1.08 &  $-$1.27 &    \nodata   & 6, 10  & 0 \\
CS~22956--050   &   $-$3.33 & 6  &  $+$1.57 &   $-$1.61 &  \nodata &  \nodata &    \nodata   & 6, 10  & 0 \\
CS~22966--057   &   $-$2.62 & 6  &  $+$2.24 &   $-$0.67 &  $-$1.27 &  $-$1.69 &    \nodata   & 6, 10  & 0 \\
CS~22968--014   &   $-$3.56 & 6  &  $+$1.46 &   \nodata &  $-$2.57 &  \nodata &    \nodata   & 6, 10  & 0 \\
CS~29491--053   &   $-$3.04 & 6  &  $+$1.83 &   $-$1.14 &  \nodata &  $-$2.94 &    \nodata   & 6, 10  & 0 \\
CS~29491--069   &   $-$2.60 & 14 &  $+$2.30 &   $-$0.17 &  $-$0.75 &  $-$0.96 &  $< +$0.35   & 14, 25 & 1 \\
CS~29495--041   &   $-$2.82 & 6  &  $+$1.93 &   $-$1.02 &  $-$2.17 &  $-$2.39 &    \nodata   & 6, 10  & 0 \\
CS~29497--004   &   $-$2.66 & 7  &  $+$2.20 &   $+$0.30 &  $-$0.38 &  $-$0.45 &    \nodata   & 7, 21  & 0 \\
CS~29516--024   &   $-$3.06 & 6  &  $+$1.74 &   $-$1.59 &  $-$2.57 &  $-$2.79 &    \nodata   & 6, 10  & 0 \\
CS~29518--051   &   $-$2.78 & 6  &  $+$2.17 &   $-$0.63 &  $-$2.17 &  \nodata &    \nodata   & 6, 10  & 0 \\
CS~30306--132   &   $-$2.42 & 16 &  \nodata &   $-$0.07 &  $-$0.78 &  $-$1.02 &  $< +$0.50   & 16     & 1 \\
CS~31078--018   &   $-$2.84 & 22 &  $+$2.13 &   $-$0.43 &  $-$1.00 &  $-$1.17 &  $< +$0.25   & 22     & 1 \\
CS~31082--001   &   $-$2.90 & 15 &  $+$1.88 &   $-$0.23 &  $-$0.62 &  $-$0.72 &    $-$0.55   & 15, 24, 31  & 3 \\
G005-001        &   $-$1.24 & 29 &  $+$3.50 &   $+$0.73 &  $-$0.09 &  $-$0.38 &  $< +$1.35   & 29, 36 & 0 \\
G009-036        &   $-$1.17 & 29 &  $+$3.14 &   $+$0.31 &  $+$0.26 &  $-$0.16 &    $+$1.25   & 29, 36 & 0 \\
G017-025        &   $-$1.54 & 29 &  $+$3.40 &   $+$0.87 &  \nodata &  \nodata &    $+$0.64   & 29, 36 & 0 \\
G023-014        &   $-$1.64 & 29 &  $+$3.05 &   $+$0.46 &  $-$0.33 &  $-$0.58 &  $< +$0.75   & 29, 36 & 1 \\
G028-043        &   $-$1.64 & 29 &  $+$3.03 &   $+$0.39 &  $-$0.22 &  $-$0.53 &    $+$0.69   & 29, 36 & 1 \\
G029-025        &   $-$1.09 & 29 &  $+$3.81 &   $+$1.14 &  $+$0.19 &  $-$0.23 &    $+$0.80   & 29, 36 & 0 \\
G040-008        &   $-$0.97 & 29 &  $+$3.90 &   $+$1.05 &  \nodata &  \nodata &    $+$1.13   & 29, 36 & 0 \\
G058-025        &   $-$1.40 & 29 &  $+$3.20 &   $+$0.76 &  $-$0.03 &  $-$0.66 &    $+$1.29   & 29, 36 & 0 \\
G059-001        &   $-$0.95 & 29 &  $+$4.04 &   $+$1.23 &  $+$0.13 &  $-$0.29 &    $+$1.64   & 29, 36 & 0 \\
G063-046        &   $-$0.90 & 29 &  $+$3.86 &   $+$1.26 &  $+$0.27 &  $-$0.05 &  $< +$1.65   & 29, 36 & 0 \\
G068-003        &   $-$0.76 & 29 &  $+$4.12 &   $+$1.35 &  $+$0.49 &  $+$0.16 &    $+$1.19   & 29, 36 & 0 \\
G074-005        &   $-$1.05 & 29 &  $+$3.56 &   $+$0.92 &  $+$0.11 &  $-$0.23 &  $< +$1.35   & 29, 36 & 0 \\
G090-025        &   $-$1.78 & 29 &  $+$2.83 &   $+$0.11 &  $-$0.51 &  $-$0.97 &  $< +$0.90   & 29, 36 & 0 \\
G095-057A       &   $-$1.22 & 29 &  $+$3.70 &   $+$1.34 &  $+$0.43 &  $-$0.23 &    $+$1.14   & 29, 36 & 0 \\
G095-057B       &   $-$1.06 & 29 &  $+$3.74 &   $+$1.30 &  \nodata &  \nodata &    $+$1.39   & 29, 36 & 0 \\
G102-020        &   $-$1.25 & 29 &  $+$3.47 &   $+$0.78 &  $-$0.02 &  $-$0.32 &    $+$0.79   & 29, 36 & 0 \\
G102-027        &   $-$0.59 & 29 &  $+$4.36 &   $+$1.62 &  $+$0.69 &  $+$0.40 &    $+$1.64   & 29, 36 & 0 \\
G113-022        &   $-$1.18 & 29 &  $+$3.64 &   $+$1.38 &  $+$0.42 &  $-$0.13 &    $+$1.19   & 29, 36 & 0 \\
G122-051        &   $-$1.43 & 29 &  $+$3.26 &   $+$0.58 &  $-$0.10 &  $-$0.27 &    $+$0.34   & 29, 36 & 3 \\
G123-009        &   $-$1.25 & 29 &  $+$3.45 &   $+$1.15 &  $+$0.21 &  $-$0.21 &    $+$1.13   & 29, 36 & 0 \\
G126-036        &   $-$1.06 & 29 &  $+$3.76 &   $+$1.59 &  $+$0.87 &  $+$0.04 &    $+$1.67   & 29, 36 & 0 \\
G126-062        &   $-$1.59 & 29 &  $+$2.86 &   $+$0.36 &  \nodata &  \nodata &  $< +$1.55   & 29, 36 & 0 \\
G140-046        &   $-$1.30 & 29 &  $+$3.65 &   $+$1.46 &  $+$0.54 &  $-$0.41 &    $+$1.19   & 29, 36 & 0 \\
G153-021        &   $-$0.70 & 29 &  $+$4.06 &   $+$1.46 &  $+$0.59 &  $+$0.34 &  $< +$1.65   & 29, 36 & 0 \\
G176-053        &   $-$1.34 & 29 &  $+$3.18 &   $+$0.63 &  $-$0.08 &  $-$0.32 &  $< +$1.05   & 29, 36 & 0 \\
G179-022        &   $-$1.35 & 29 &  $+$3.34 &   $+$0.79 &  $+$0.02 &  $-$0.22 &    $+$0.39   & 29, 36 & 0 \\
G180-024        &   $-$1.34 & 29 &  $+$3.23 &   $+$0.72 &  $-$0.19 &  $-$0.58 &  $< +$1.65   & 29, 36 & 0 \\
G188-022        &   $-$1.52 & 29 &  $+$3.24 &   $+$0.94 &  $-$0.12 &  $-$0.60 &  $< +$1.30   & 29, 36 & 0 \\
G191-055        &   $-$1.63 & 29 &  $+$2.78 &   $+$0.23 &  $-$0.41 &  $-$0.89 &  $< +$1.10   & 29, 36 & 0 \\
G192-043        &   $-$1.50 & 29 &  $+$3.12 &   $+$0.71 &  $-$0.02 &  $-$0.25 &  $< +$1.60   & 29, 36 & 0 \\
G221-007        &   $-$0.98 & 29 &  $+$3.65 &   $+$1.02 &  $+$0.28 &  $-$0.11 &    $+$1.69   & 29, 36 & 0 \\
HD~2665         &   $-$1.99 & 29 &  $+$2.49 &   $-$0.42 &  $-$0.92 &  $-$1.15 &  $< +$0.30   & 29, 36 & 1 \\
HD~2796         &   $-$2.47 & 6  &  $+$2.37 &   $-$0.51 &  $-$1.47 &  $-$1.84 &    \nodata   & 6, 10  & 0 \\
HD~8724         &   $-$1.91 & 29 &  $+$2.90 &   $+$0.25 &  $-$0.49 &  $-$0.86 &    $+$0.00   & 29, 36 & 3 \\
HD~3008         &   $-$2.08 & 29 &  $+$2.52 &   $-$0.30 &  $-$0.79 &  $-$1.09 &    $-$0.56   & 29, 36 & 3 \\
HD~6268         &   $-$2.42 & 9  &  \nodata &   $-$0.38 &  $-$1.05 &  $-$1.37 &  $< +$0.08   & 16, 25 & 1 \\
HD~6755         &   $-$1.68 & 29 &  $+$2.87 &   $+$0.31 &  $-$0.29 &  $-$0.50 &    $+$0.32   & 29, 36 & 3 \\
HD~6833         &   $-$0.85 & 29 &  $+$3.77 &   $+$1.22 &  $+$0.34 &  $+$0.10 &    $+$1.69   & 29, 36 & 0 \\
HD~13979        &   $-$2.92 & 27 &  $+$1.79 &   $-$1.30 &  $-$2.30 &  $-$2.59 &  $< +$0.45   & 27     & 0 \\
HD~21581        &   $-$1.71 & 29 &  $+$2.99 &   $+$0.44 &  $-$0.42 &  $-$0.81 &  $< +$0.50   & 29, 36 & 1 \\
HD~23798        &   $-$2.26 & 29 &  $+$2.39 &   $-$0.19 &  $-$1.01 &  $-$1.36 &    $-$0.21   & 29, 36 & 1 \\
HD~25329        &   $-$1.67 & 29 &  $+$3.00 &   $+$0.67 &  $-$0.05 &  \nodata &    \nodata   & 29, 36 & 0 \\
HD~25532        &   $-$1.34 & 29 &  $+$3.52 &   $+$1.05 &  $-$0.13 &  $-$0.64 &  $< +$1.15   & 29, 36 & 0 \\
HD~26297        &   $-$1.98 & 29 &  $+$3.07 &   $+$0.40 &  $-$0.85 &  $-$1.22 &    $-$0.11   & 29, 36 & 1 \\
HD~29574        &   $-$2.00 & 29 &  $+$2.67 &   $+$0.19 &  $-$0.37 &  $-$0.63 &    $-$0.06   & 29, 36 & 3 \\
HD~37828        &   $-$1.62 & 29 &  $+$3.18 &   $+$0.70 &  $-$0.12 &  $-$0.53 &    $+$0.77   & 29, 36 & 1 \\
HD~44007        &   $-$1.72 & 29 &  $+$2.83 &   $+$0.26 &  $-$0.51 &  $-$0.94 &    $+$0.31   & 29, 36 & 1 \\
HD~63791        &   $-$1.90 & 29 &  $+$2.87 &   $+$0.23 &  $-$0.55 &  $-$0.92 &    $+$0.22   & 29, 36 & 1 \\
HD~74462        &   $-$1.52 & 29 &  $+$3.12 &   $+$0.51 &  $-$0.17 &  $-$0.39 &    $+$0.49   & 29, 36 & 3 \\
HD~82590        &   $-$1.32 & 29 &  $+$3.03 &   $+$0.71 &  $-$0.11 &  $-$0.46 &  $< +$1.60   & 29, 36 & 0 \\
HD~85773        &   $-$2.62 & 29 &  $+$2.56 &   $-$0.93 &  $-$1.56 &  $-$1.84 &  $< -$0.15   & 29, 36 & 1 \\
HD~88609        &   $-$3.07 & 18 &  $+$1.77 &   $-$0.97 &  $-$2.75 &  $-$2.89 &    \nodata   & 25, 36 & 3 \\
HD~101063       &   $-$1.33 & 29 &  $+$3.30 &   $+$0.85 &  $+$0.21 &  $+$0.00 &  $< +$1.10   & 29, 36 & 0 \\
HD~103036       &   $-$2.04 & 29 &  $+$2.80 &   $+$0.13 &  $-$0.62 &  $-$1.09 &  $< +$0.10   & 29, 36 & 1 \\
HD~103545       &   $-$2.45 & 29 &  $+$2.20 &   $-$0.50 &  $-$1.18 &  $-$1.56 &  $< -$0.05   & 29, 36 & 1 \\
HD~105546       &   $-$1.48 & 29 &  $+$3.29 &   $+$0.74 &  $-$0.13 &  $-$0.56 &  $< +$0.80   & 29, 36 & 1 \\
HD~105755       &   $-$0.83 & 29 &  $+$3.98 &   $+$1.18 &  $+$0.33 &  $+$0.02 &    $+$1.43   & 29, 36 & 0 \\
HD~106516       &   $-$0.81 & 29 &  $+$3.94 &   $+$1.28 &  $+$0.31 &  $-$0.04 &    $+$1.56   & 29, 36 & 0 \\
HD~107752       &   $-$2.78 & 29 &  $+$1.93 &   $-$0.90 &  $-$1.59 &  $-$1.99 &  $< +$0.30   & 29, 36 & 0 \\
HD~108317       &   $-$2.18 & 29 &  $+$2.40 &   $-$0.39 &  $-$1.01 &  $-$1.32 &    $+$0.17   & 25, 36 & 1 \\
HD~108577       &   $-$2.38 & 20 &  $+$2.56 &   $-$0.52 &  $-$1.24 &  $-$1.48 &    \nodata   & 20     & 0 \\
HD~110184       &   $-$2.72 & 29 &  $+$2.14 &   $-$0.65 &  $-$1.47 &  $-$1.71 &  $< -$0.20   & 29, 36 & 1 \\
HD~115444       &   $-$2.90 & 29 &  $+$1.90 &   $-$0.82 &  $-$1.42 &  $-$1.64 &  $< -$0.45   & 25, 36 & 1 \\
HD~119516       &   $-$2.26 & 26 &  $+$2.32 &   $-$0.43 &  $-$1.08 &  $-$1.43 &  $< +$1.52   & 26     & 2 \\
HD~121135       &   $-$1.54 & 29 &  $+$3.37 &   $+$0.69 &  $-$0.33 &  $-$0.70 &    $+$0.38   & 29, 36 & 1 \\
HD~122563       &   $-$2.77 & 17 &  $+$1.97 &   $-$0.92 &  $-$2.40 &  $-$2.75 &  $< -$0.42   & 25, 36 & 3 \\
HD~122956       &   $-$1.95 & 29 &  $+$2.87 &   $+$0.16 &  $-$0.48 &  $-$0.79 &    $-$0.13   & 29, 36 & 3 \\
HD~124358       &   $-$1.91 & 29 &  $+$2.64 &   $-$0.22 &  $-$0.68 &  $-$0.94 &  $< +$0.45   & 29, 36 & 1 \\
HD~126587       &   $-$2.93 & 9  &  \nodata &   $-$0.53 &  $-$1.75 &  $-$1.97 &  $< -$0.38   & 16, 25 & 1 \\
HD~128279       &   $-$2.51 & 26 &  $+$2.15 &   $-$1.04 &  $-$1.77 &  $-$2.27 &  $< +$1.05   & 26     & 2 \\
HD~132475       &   $-$1.86 & 29 &  $+$2.96 &   $+$0.56 &  $-$0.38 &  $-$0.92 &  $< +$1.00   & 29, 36 & 0 \\
HD~135148       &   $-$2.17 & 29 &  $+$2.58 &   $-$0.25 &  $-$0.76 &  $-$0.95 &    $-$0.17   & 29, 36 & 3 \\
HD~141531       &   $-$1.79 & 29 &  $+$2.79 &   $+$0.11 &  $-$0.45 &  $-$0.72 &    $+$0.20   & 29, 36 & 1 \\
HD~166161       &   $-$1.23 & 29 &  $+$3.46 &   $+$1.11 &  $+$0.23 &  $-$0.48 &    $+$0.84   & 29, 36 & 0 \\
HD~171496       &   $-$0.67 & 29 &  $+$4.11 &   $+$1.40 &  $+$0.51 &  $+$0.11 &    $+$1.41   & 29, 36 & 0 \\
HD~175305       &   $-$1.73 & 26 &  $+$2.96 &   $+$0.30 &  $-$0.60 &  $-$0.89 &    $-$0.28   & 25, 26 & 3 \\
HD~184266       &   $-$1.43 & 29 &  $+$3.19 &   $+$0.69 &  $-$0.11 &  $-$0.43 &  $< +$1.60   & 29, 36 & 0 \\
HD~186478       &   $-$2.56 & 29 &  $+$2.23 &   $-$0.45 &  $-$1.32 &  $-$1.53 &  $< -$0.26   & 25, 36 & 1 \\
HD~187111       &   $-$1.97 & 29 &  $+$2.84 &   $+$0.16 &  $-$0.57 &  $-$0.88 &    $-$0.10   & 29, 36 & 3 \\
HD~188510       &   $-$1.32 & 29 &  $+$3.01 &   $+$0.44 &  $-$0.14 &  $-$0.52 &  $< +$1.20   & 29, 36 & 0 \\
HD~193901       &   $-$1.08 & 29 &  $+$3.36 &   $+$0.83 &  $+$0.19 &  $-$0.10 &  $< +$1.45   & 29, 36 & 0 \\
HD~194598       &   $-$1.08 & 29 &  $+$3.40 &   $+$0.90 &  $+$0.08 &  $-$0.28 &  $< +$1.40   & 29, 36 & 0 \\
HD~201891       &   $-$1.09 & 29 &  $+$3.55 &   $+$0.88 &  $+$0.12 &  $-$0.22 &    $+$1.25   & 29, 36 & 0 \\
HD~204543       &   $-$1.87 & 29 &  \nodata &   $+$0.12 &  $-$0.63 &  $-$1.05 &    $+$0.05   & 5, 25  & 1 \\
HD~206739       &   $-$1.72 & 29 &  $+$2.98 &   $+$0.36 &  $-$0.32 &  $-$0.62 &    $+$0.38   & 29, 36 & 1 \\
HD~210295       &   $-$1.46 & 29 &  $+$3.37 &   $+$0.85 &  $-$0.10 &  $-$0.34 &    $+$0.72   & 29, 36 & 1 \\
HD~214362       &   $-$1.87 & 29 &  $+$2.71 &   $+$0.32 &  $-$0.48 &  $-$0.82 &  $< +$1.00   & 29, 36 & 0 \\
HD~214925       &   $-$2.08 & 2  &  \nodata &   \nodata &  $-$0.86 &  $-$1.09 &    $-$0.50   & 2      & 3 \\
HD~216143       &   $-$2.32 & 2  &  $+$2.57 &   $-$0.12 &  $-$1.21 &  $-$1.24 &  $< -$0.10   & 2, 13  & 1 \\
HD~218857       &   $-$1.90 & 29 &  $+$2.64 &   $-$0.19 &  $-$1.16 &  $-$1.42 &  $< +$0.55   & 29, 36 & 0 \\
HD~220838       &   $-$1.80 & 2  &  $+$999. &   $+$0.47 &  $-$0.76 &  $-$0.93 &    $+$0.05   & 2, 4   & 1 \\
HD~221170       &   $-$2.16 & 19 &  $+$2.51 &   $-$0.08 &  $-$0.73 &  $-$0.86 &    $-$0.09   & 19     & 3 \\
HD~233666       &   $-$1.79 & 29 &  $+$2.88 &   $+$0.22 &  $-$0.68 &  $-$1.03 &  $< +$0.40   & 29, 36 & 1 \\
HD~235766       &   $-$1.93 & 2  &  \nodata &   \nodata &  $-$0.60 &  $-$0.86 &    $+$0.10   & 2      & 1 \\
HD~237846       &   $-$3.29 & 26 &  $+$1.69 &   $-$1.56 &  \nodata &  $-$3.10 &    $+$0.29   & 26     & 2 \\
HE~0430$-$4901  &   $-$2.72 & 3  &  \nodata &   $-$0.45 &  \nodata &  $-$1.05 &    \nodata   & 3      & 0 \\
HE~0432$-$0923  &   $-$3.19 & 3  &  \nodata &   $-$0.44 &  \nodata &  $-$1.43 &    \nodata   & 3      & 0 \\
HE~1127$-$1143  &   $-$2.73 & 3  &  \nodata &   $-$0.27 &  \nodata &  $-$1.14 &    \nodata   & 3      & 0 \\
HE~1219$-$0312  &   $-$2.97 & 14 &  $+$1.78 &   $-$0.40 &  $-$0.75 &  $-$0.98 &  $< +$0.53   & 14, 25 & 1 \\
HE~1523$-$0901  &   $-$2.95 & 11 &  \nodata &   $-$0.27 &  $-$0.63 &  $-$0.62 &  $< -$0.20   & 11, 12 & 3 \\
HE~2224$+$0143  &   $-$2.58 & 3  &  $+$2.29 &   $-$2.22 &  $-$0.77 &  $-$1.02 &    \nodata   & 3      & 0 \\
HE~2327$-$5642  &   $-$2.79 & 3  &  $+$1.83 &   $-$0.69 &  $-$1.10 &  $-$1.29 &    \nodata   & 23     & 0 \\
M5~IV--81       &   $-$1.28 & 34 &  $+$3.21 &   $+$1.15 &  $+$0.11 &  $-$0.31 &    $+$0.35   & 34, 35 & 0 \\
M5~IV--82       &   $-$1.33 & 34 &  $+$3.21 &   $+$1.00 &  $+$0.11 &  $-$0.23 &    $+$0.25   & 34, 35 & 0 \\
M13~L598        &   $-$1.56 & 33 &  \nodata &   $+$0.55 &  $-$0.34 &  $-$0.58 &    $+$0.09   & 33     & 3 \\
M13~L629        &   $-$1.63 & 33 &  \nodata &   $+$0.63 &  $-$0.35 &  $-$0.61 &    $+$0.12   & 33     & 3 \\
M13~L70         &   $-$1.59 & 33 &  \nodata &   $+$0.50 &  $-$0.23 &  $-$0.58 &    $+$0.09   & 33     & 3 \\
M13~L973        &   $-$1.61 & 33 &  \nodata &   $+$0.55 &  $-$0.27 &  $-$0.51 &    $-$0.01   & 33     & 3 \\
M15~K341        &   $-$2.54 & 32 &  $+$2.04 &   $-$0.49 &  $-$1.28 &  $-$1.52 &    \nodata   & 32     & 0 \\
M15~K462        &   $-$2.55 & 32 &  $+$2.00 &   $-$0.41 &  $-$1.03 &  $-$1.20 &    \nodata   & 32     & 0 \\
M15~K583        &   $-$2.58 & 32 &  $+$1.99 &   $-$0.63 &  $-$1.52 &  $-$1.80 &    \nodata   & 32     & 0 \\
M92~VII-18      &   $-$2.29 & 20 &  \nodata &   $-$0.20 &  $-$1.29 &  $-$1.45 &    \nodata   & 20     & 0 \\
NGC~6752~B702   &   $-$1.58 & 33 &  \nodata &   $+$0.67 &  $-$0.39 &  $-$0.78 &    $+$0.27   & 33     & 1 \\
NGC~6752~B708   &   $-$1.63 & 33 &  \nodata &   $+$0.62 &  $-$0.50 &  $-$0.83 &    $+$0.17   & 33     & 1 \\
NGC~6752~PD1    &   $-$1.62 & 33 &  \nodata &   $+$0.66 &  $-$0.45 &  $-$0.78 &    $+$0.03   & 33     & 1 \\
NGC~6752~B1630  &   $-$1.60 & 33 &  \nodata &   $+$0.65 &  $-$0.45 &  $-$0.74 &    $+$0.25   & 33     & 1 \\
NGC~6752~B3589  &   $-$1.59 & 33 &  \nodata &   $+$0.72 &  $-$0.41 &  $-$0.72 &    $+$0.18   & 33     & 1 \\
UMi~COS82       &   $-$1.42 & 1  &  $+$2.82 &   $+$1.22 &  $+$0.52 &  $+$0.34 &    \nodata   & 1, 28  & 0 \\
\enddata
\tablenotetext{a}{Classifications:
(1) log~(Pb/Eu)~$< +$1.8;
(2) member of the stellar stream analyzed by \citet{roederer10a};
(3) log~(Pb/Eu)~$< +0.9$, as well as \mbox{HD~88609} and \mbox{HD~122563};
(0) none of 1--3 or [Fe/H]~$\geq -$1.4.
}
\tablerefs{
(1) \citet{aoki07};
(2) \citet{aoki08};
(3) \citet{barklem05};
(4) \citet{burris00};
(5) \citet{burris09};
(6) \citet{cayrel04};
(7) \citet{christlieb04};
(8) \citet{cowan02};
(9) \citet{cowan05};
(10) \citet{francois07};
(11) \citet{frebel07};
(12) A.\ Frebel (2009, private communication);
(13) \citet{fulbright00};
(14) \citet{hayek09};
(15) \citet{hill02};
(16) \citet{honda04};
(17) \citet{honda06};
(18) \citet{honda07};
(19) \citet{ivans06};
(20) \citet{johnson02};
(21) \citet{jonsell06};
(22) \citet{lai08};
(23) \citet{mashonkina10};
(24) \citet{plez04};
(25) \citet{roederer09b};
(26) \citet{roederer10a};
(27) I.\ Roederer et al.\ (in preparation);
(28) \citet{sadakane04};
(29) \citet{simmerer04};
(30) \citet{sneden03};
(31) \citet{sneden09};
(32) \citet{sobeck10};
(33) \citet{yong06};
(34) \citet{yong08a};
(35) \citet{yong08b};
(36) this study.
}
\end{deluxetable}

\vspace*{1.5in}

\begin{deluxetable}{cccccc}
\tablecaption{Surface Composition of AGB Models
\label{agbtab}}
\tablewidth{0pt}
\tabletypesize{\scriptsize}
\tablehead{
\colhead{$M_{0}$ ($M_{\odot}$)} &
\colhead{[La/Eu]} &
\colhead{[Pb/Eu]} &
\colhead{[Pb/Fe]} &
\colhead{[Eu/Fe]} &
\colhead{$^{13}$C Pocket Extent in Mass} }
\startdata
\hline
\multicolumn{6}{c}{[Fe/H]~$= -$1.4} \\
\hline
1.25& 0.61 & 1.66 & 1.81 & 0.14 & 2$\times10^{-3} M_{\odot}$ \\
2.5 & 0.90 & 1.87 & 2.85 & 0.98 & 2$\times10^{-3} M_{\odot}$ \\
3.5 & 0.84 & 1.82 & 2.54 & 0.72 & 1$\times10^{-3} M_{\odot}$ \\
\hline
\multicolumn{6}{c}{[Fe/H]~$= -$2.3} \\
\hline
1.0 & 1.00 & 1.61 & 2.66 & 1.04 & 2$\times10^{-3} M_{\odot}$ \\
1.5 & 0.99 & 1.65 & 3.10 & 1.45 & 2$\times10^{-3} M_{\odot}$ \\
    & 0.98 & 1.72 & 3.17 & 1.46 & 4$\times10^{-3} M_{\odot}$ \\
2.0 & 0.95 & 1.70 & 3.18 & 1.48 & 2$\times10^{-3} M_{\odot}$ \\
    & 1.06 & 1.93 & 3.56 & 1.62 & 4$\times10^{-3} M_{\odot}$ \\
4.5 & 0.89 & 0.37 & 0.57 & 0.20 & no pocket \\
5.0 & 0.92 & 0.47 & 0.73 & 0.26 & no pocket \\
6.0 & 1.04 & 0.58 & 0.99 & 0.42 & no pocket \\
\enddata
\tablecomments{
All models assume a scaled-solar initial composition
\citep{asplund09}.}
\end{deluxetable}

\end{center}

\end{document}